\documentclass[aapm,mph,reprint,superscriptaddress,nofootinbib]{revtex4}
\usepackage[italian]{varioref}
\usepackage{lipsum}
\usepackage{amssymb,amsmath,physics,booktabs,subfig,mathtools,xfrac}
\usepackage[]{graphicx}
\usepackage{ragged2e}
\usepackage{float}
\usepackage{caption}
\usepackage{gensymb,mathtools,amsfonts,amsthm,mathrsfs,slashed,relsize}
\usepackage[output-decimal-marker={.}]{siunitx}
\usepackage{xcolor,tcolorbox}
\usepackage{appendix}
\usepackage{dcolumn}
\usepackage{bm}
\usepackage{comment}
\usepackage{xcolor}
\usepackage[mathlines]{lineno}
\usepackage[colorlinks]{hyperref}
\modulolinenumbers[5]

\newcommand{\versor}[1]{\mathbf{\hat{#1}}}

\hypersetup{hidelinks}

\begin{document}
	
\preprint{AAPM/123-QED}
	
\title[]{Probing Axions through Tomography of Anisotropic Cosmic Birefringence}
	
\author{Alessandro Greco}
\affiliation{Dipartimento di Fisica e Astronomia ``Galileo Galilei", Universit\`a degli Studi di Padova, via Marzolo 8, I-35131, Padova, Italy}
\affiliation{INFN, Sezione di Padova, via Marzolo 8, I-35131, Padova, Italy}
	
\author{Nicola Bartolo}
\affiliation{Dipartimento di Fisica e Astronomia ``Galileo Galilei", Universit\`a degli Studi di Padova, via Marzolo 8, I-35131, Padova, Italy}
\affiliation{INFN, Sezione di Padova, via Marzolo 8, I-35131, Padova, Italy}
\affiliation{INAF - Osservatorio Astronomico di Padova, vicolo dell' Osservatorio 5, I-35122 Padova, Italy}
	
\author{Alessandro Gruppuso}
\affiliation{Istituto Nazionale di Astrofisica - Osservatorio di Astrofisica e Scienza dello Spazio di Bologna, via Gobetti 101, I-40129 Bologna, Italy}
\affiliation{INFN, Sezione di Bologna, viale Berti Pichat 6/2, I-40127 Bologna, Italy}
\affiliation{Dipartimento di Fisica e Scienze della Terra, Universit\`a degli Studi di Ferrara, via Saragat 1,I-44122, Ferrara, Italy}
	
\date{\today}
	
\begin{abstract}
Cosmic birefringence is the in-vacuo rotation of the linear polarization plane experienced by photons of the Cosmic Microwave Background (CMB) radiation when theoretically well-motivated parity-violating extensions of Maxwell electromagnetism are considered. If the angle parametrizing such a rotation is dependent on the photon's direction, then this phenomenon is called Anisotropic Cosmic Birefringence (ACB). In this paper, we perform for the first time a tomographic treatment of the ACB, by considering photons emitted both at the recombination and reionization epoch. This allows one to extract additional and complementary information about the physical source of cosmic birefringence with respect to the isotropic case. We focus here on the case of an axion-like field $\chi$, whose coupling with the electromagnetic sector induces such a phenomenon, by using an analytical and numerical approach (which involves a modification of the \texttt{CLASS} code). We find that the anisotropic component of cosmic birefringence exhibits a peculiar behavior: an increase of the axion mass implies an enhancement of the anisotropic amplitude, allowing to probe a wider range of masses with respect to the purely isotropic case. Moreover, we show that at large angular scales, the interplay between the reionization and recombination contributions to ACB is sensitive to the axion mass, so that at sufficiently low multipoles, for sufficiently light masses, the reionization contribution overtakes the recombination one, making the tomographic approach to cosmic birefringence a promising tool for investigating the properties of this axion-like field.
\end{abstract}
	
\keywords{Cosmic Birefringence, Parity-Violation, CMB Anisotropies.}
\maketitle
	
\section{\label{sec:Intro}Introduction}
Observing CMB photons gives the opportunity to test fundamental physics, and in particular modifications of the electromagnetic theory \cite{lue1999cosmological,caloni2023probing}. A well known extension of the standard formulation is obtained by adding an extra-term in the Maxwell Lagrangian density which encodes a Chern-Simons interaction between photons and a new scalar field $\chi$ \cite{carroll1990limits}:
\begin{equation}
\label{eqn:Chern-Simons}
\mathcal{S}=\int\mathrm{d}^4x\,\sqrt{-g}\left[-\frac{1}{2}g^{\mu\nu}\partial_{\mu}\chi\partial_{\nu}\chi-V(\chi)-\frac{1}{4}F_{\mu\nu}F^{\mu\nu}-\frac{\lambda}{4f}\chi F_{\mu\nu}\tilde{F}^{\mu\nu}\right],
\end{equation}
where $g$ is the determinant of the metric tensor, $V(\chi)$ is the scalar potential of the field $\chi$, $\lambda$ is a dimensionless coupling constant, $f$ is a decay constant with dimension of energy, and $\tilde{F}^{\mu\nu}\equiv\epsilon^{\mu\nu\rho\sigma}F_{\rho\sigma}/2$
is the Hodge dual of the Maxwell tensor $F_{\mu\nu}$. Since $F_{\mu\nu}\tilde{F}^{\mu\nu}$ violates parity symmetry, the field $\chi$ has to be a pseudo-scalar in order to make the Lagrangian density parity-invariant, and for this reason such a field can be identified as
an axion-like field \cite{takahashi2021kilobyte,choi2021cosmic,nakagawa2021cosmic,gasparotto2022cosmic,kitajima2022power,gonzalez2022stability,cao2022non}. Physically, the existence of such a field can have various motivations, including predictions of string theory \cite{arvanitaki2010string,hlozek2015search,marsh2016axion,kim2021cosmic,jain2021cmb,lin2022consistency}, so that it is possible to use cosmology for investigating physics beyond the standard model. In the literature $\chi$ has been proposed as a candidate for dark matter \cite{preskill1983cosmology,abbott1983cosmological,dine1983not,liu2017axion,obata2022implications,zhou2023cosmic}, quintessence models \cite{caldwell2011cross,fujita2021detection}, or for early dark energy in the form of a pseudo Nambu-Goldstone boson to explain the Hubble tension \cite{poulin2018cosmological,poulin2019early,capparelli2020cosmic,rezazadeh2022cascading,kamionkowski2022the}, making CMB physics a natural framework to unveil the dark sector of the Universe.

The main feature of the Chern-Simons modified theory of electromagnetism, is that it causes a rotation of the linear polarization plane of the electromagnetic waves. This can be mathematically described as the rotation of the Stokes parameters $Q$ and $U$ experienced by photons when they propagate \cite{liu2006effect,li2008cosmological,namikawa2021cmb,komatsu2022new}:
\begin{equation}
\label{eqn:rotation}
[Q\pm iU]\mapsto[Q\pm iU]e^{\pm2i\alpha}.
\end{equation}
Since the standard electromagnetic theory has historically robust scientific evidences of being able to describe the light's propagation, it is reasonable to expect the quantity $\alpha$, where $\alpha$ is the \textit{birefringence angle}, to be small. 

As can be understood by looking at Eq.~\eqref{eqn:rotation}, the presence of cosmic birefringence effects implies a modification of the observed CMB power spectra and in particular provides a non-vanishing value for the parity-breaking cross-correlations $TB$ and $EB$, see e.g. \cite{feng2006searching},
\begin{align}
\label{eqn:TEobs}
C^{TE}_{\ell,\text{obs}}&= C_{\ell}^{TE}\cos2\alpha_0,\\	
C^{TB}_{\ell,\text{obs}}&= C^{TE}_{\ell}\sin2\alpha_0,\\
C^{EE}_{\ell,\text{obs}}&= C_{\ell}^{EE}\cos^22\alpha_0+C_{\ell}^{BB}\sin^22\alpha_0,\\
C^{EB}_{\ell,\text{obs}}&=\frac{1}{2}\left(C_{\ell}^{EE}-C_{\ell}^{BB}\right)\sin4\alpha_0,\\
\label{eqn:BBobs}
C^{BB}_{\ell,\text{obs}}&=C_{\ell}^{EE}\sin^22\alpha_0+C_{\ell}^{BB}\cos^22\alpha_0\, ,
\end{align}
where $\alpha_0$ is the \textit{isotropic birefringence angle}, which is related to the field $\chi$ in the following way (see e.g. \cite{komatsu2022new} and Refs. therein):
\begin{equation}
\label{eqn:iso_angle}
\alpha_0=\frac{\lambda}{2f}\left[\chi_0(\tau_0)-\chi_0(\tau_{\text{reco}})\right],
\end{equation}
where $\tau_0$ and $\tau_{\text{reco}}$ are respectively, the conformal time today and at the recombination epoch, when photons were mainly emitted. Eq.~\eqref{eqn:iso_angle} holds true only if $\chi=\chi_0(\tau)$ is a homogeneous field and by working in the sudden recombination approximation: we will see very soon how the situation changes if we relax these assumptions. As can be easily seen by looking at Eqs.~\eqref{eqn:TEobs}-\eqref{eqn:BBobs}, if we set $\alpha_{0}=0$ we recover the standard $\Lambda$CDM results.

Eq.~\eqref{eqn:iso_angle} shows that birefringence is a propagation effect and therefore larger is the path of the photon, larger will be the probability for the field $\chi$ to evolve enough to produce an appreciable value for $\alpha_{0}$. This is the reason why the most promising observations of such a phenomenon come from cosmology: since CMB photons represent the oldest source of electromagnetic radiation in the Universe, they have traveled the longest possible path.

Constraining parity-violation from CMB data is a well known historical effort for cosmologists\footnote{Other parity-violating signatures which can be constrained with CMB observations arise in the tensor sector when considering chiral primordial gravitational waves  \cite{alexander2005birefringent,contaldi2008anomalous,alexander2009chern,bartolo2015parity,bartolo2015non,qiao2019waveform,qiao2020polarized}. However, since constraining this kind of parity-breaking effects by looking at the CMB power spectra holds to be extremely hard and challenging \cite{gerbino2016testing}, in the last years the possibility to test gravitational waves' chirality through the three-point angular correlation function of CMB has received growing interest (see e.g.\cite{maldacena2011graviton,kamionkowski2011odd,shiraishi2013bispectrum,shiraishi2013gravitons,shiraishi2015observed,meerburg2016cmb,bartolo2017parity,bartolo2019measuring,duivenvoorden2020cmb,bartolo2021tensor}).}, and intriguing signals seem to come from what concerns cosmic birefringence \cite{aghanim2016planck,fujita2022can}. Indeed, in the last years there has been an increasing amount of important observational constraints on isotropic cosmic birefringence: these results are collected in Tab.~\ref{tab:iso_constraints}. By using \textit{Planck} maps from the third public release (PR3), it has been possible to extract a promising measurement of a non-vanishing birefringence angle \cite{minami2020new}, and then such a treatment has been also extended to PR4 \cite{diego2022cosmic}. Similar results come by a joint analysis of polarization data from the space missions WMAP and \textit{Planck} \cite{eskilt2022improved}.

Although there exists the possibility that this effect is caused by interstellar dust emission \cite{clark2021origin,cukierman2022magnetic,eskilt2022frequency,vacher2022frequency}, confirming these detections by observations with higher statistical significance in the future might have a profound implication for fundamental physics \cite{diego2022robustness,monelli2022impact,jost2022characterising}. Results reported in Tab.~\ref{tab:iso_constraints} have been possible thanks to a new technique which takes also into account information present in the Galactic foreground emission. Indeed, if one relied only on the CMB power spectra, it would not be possible to distinguish between $\alpha_{0}$ and a miscalibration of the instrumental polarization angle, say $\vartheta$; thus, $\alpha_{0}$ and $\vartheta$ would be degenerate \cite{de2022determination,miller2009cmb,jarosik2011seven,keating2012self}. However, since $\alpha_{0}$ is proportional to the difference between the value of the axion field today and the one at the emission time of the photon, one can reasonably expect that the polarized emission from our Galaxy is only negligibly affected by cosmic birefringence. Therefore, it is possible to use this property to break the degeneracy between $\alpha_{0}$ and $\vartheta$ in order to isolate the two different rotation angles \cite{minami2019simultaneous}.

\begin{table}[h]
\caption{\label{tab:iso_constraints}Most recent measurements of the isotropic birefringence angle coming from CMB observations \cite{minami2020new,diego2022cosmic,eskilt2022improved}.}
\begin{tabular}{|c|c|c|}
\hline
\textbf{PR3} & \textbf{PR4} & \textbf{\textit{Planck} + WMAP} \\
\hline
$\alpha_0=(0.35\pm0.14)^{\circ}$ & $\alpha_0=(0.30\pm0.11)^{\circ}$ & $\alpha_0={0.342}^{\circ+0.094^{\circ}}_{\,\,\,-0.091^{\circ}}$ \\
\hline
\end{tabular}
\end{table}

As previously mentioned, Eq.~\eqref{eqn:iso_angle} is valid only if we assume that photons of the CMB were all emitted at the recombination time. However, they are instead statistically distributed over the photon visibility function, and a small but not negligible amount of them was also emitted at the reionization epoch. Since the reionization CMB polarization signal arises from a much lower redshift than the recombination CMB polarization signal, it is reasonable to expect that the birefringence angle for reionization polarization $\alpha_{0,\,\text{reio}}$ may in general be different from the birefringence angle for recombination polarization $\alpha_{0,\,\text{reco}}$ \cite{sherwin2021cosmic,nakatsuka2022cosmic,lee2022probing,galaverni2023redshift}. How much the two angles differ from each other strongly depends on the background evolution of the field $\chi_0(\tau)$, since the formula in Eq.~\eqref{eqn:iso_angle} can be generalized to
\begin{equation}
\label{eqn:gen_iso_angle}
\alpha_{0}(\tau)=\frac{\lambda}{2f}\left[\chi_0(\tau_0)-\chi_0(\tau)\right],
\end{equation}
where $\tau$ is the conformal time at which the photon has been emitted. In other words, $\alpha_{0,\,\text{reio}}$ is different from $\alpha_{0,\,\text{reco}}$ only if the field $\chi_0$ has been able to evolve in such a way to get a different value at reionization with respect to the one at recombination. Moreover, in general non-vanishing birefringence angle is generated only if the coupling parameter $\lambda/f$ is different from zero and if the value of the axion-like field today $\chi_0(\tau_0)$ is sufficiently different from the value at the emission time of the photon, i.e. $\chi_{0}(\tau)$.

Since cosmic birefringence is strictly related to the physical evolution of the field $\chi$, the aim of this paper is to study separately the contributions from the two epochs (recombination and reionization), and to perform a tomographic analysis of cosmic birefringence, which can be used to infer important information about the underlying physics of $\chi$. Currently, this kind of approach has been limited to isotropic birefringence \cite{sherwin2021cosmic,nakatsuka2022cosmic,lee2022probing}, and the purpose of our work is to extend such a treatment also to \textit{anisotropic} birefringence. Indeed, as we are going to show in the next sections, anisotropic cosmic birefringence (ACB) encodes additional information on the physical model that describes the field $\chi$, and, in analogy with the isotropic case, a tomographic approach is a well suited way to probe the underlying axion model. More specifically, in this paper we show that, differently from the isotropic case, a large mass of the axion field does not prevent the possibility to induce cosmic birefringence, making the anisotropic signal able to probe a wider range of masses for the axion field with respect to the purely isotropic regime.

The structure of the paper is organized as follows. In Sec.~\ref{sec:chi_0} we study the time evolution of $\chi_0(\tau)$ to see how it is affected by the value of the axion mass. In Sec.~\ref{sec:Aniso} we briefly review anisotropic cosmic birefringence, and we compute the angular power spectrum of the birefringence angle and its two-point cross-correlation with CMB temperature and $E$ polarization modes for different values of the axion mass. In Sec.~\ref{sec:Tomo} we perform a tomographic analysis, by deriving some general formulas in order to show what is the impact of the two birefringence signals (from recombination and reionization epochs, respectively) on the CMB angular power spectra. Sec.~\ref{sec:End} is dedicated to the conclusions and discussions. Some details about the analytical derivation of the expressions for the birefringent CMB angular power spectra can be found in App.~\ref{App:CMB}.

\section{\label{sec:chi_0}Background Evolution of the Axion-like Field}
As can be easily seen by looking at Eq.~\eqref{eqn:gen_iso_angle}, in order to understand what is the interplay between the recombination and reionization contributions to isotropic cosmic birefringence, we need to study the background dynamics of the axion-like field $\chi_0(\tau)$.

\begin{figure}
\centering
\subfloat[][\label{fig:chi_background}Evolution of the ratio between the homogeneous scalar field $\chi_0$ and its initial value $\chi_{0}^{\text{ini}}$.]
{\includegraphics[width=.75\textwidth]{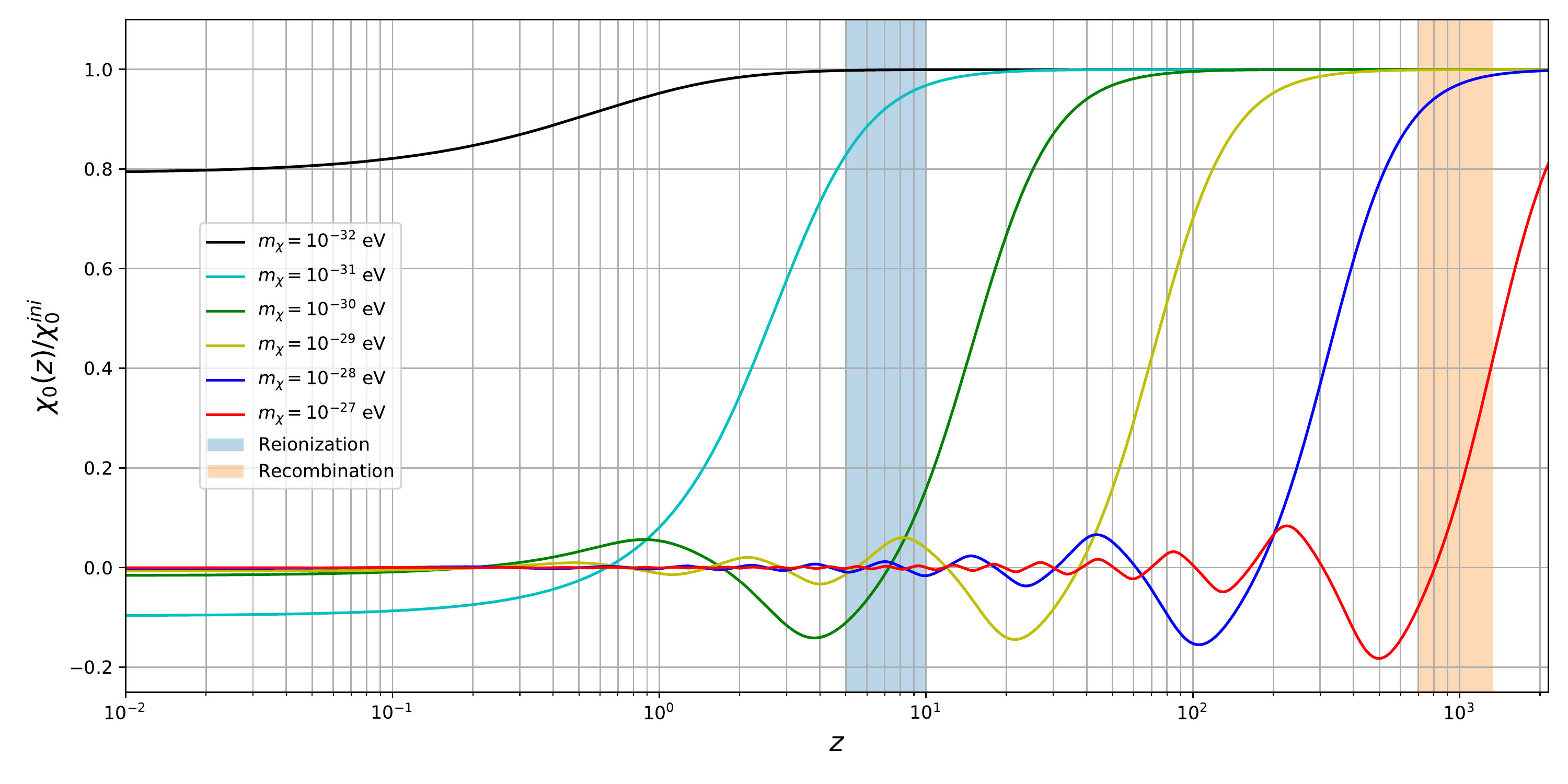}} \\
\subfloat[][\label{fig:w_chi}Evolution of the equation of state for the homogeneous scalar field $\chi_0$.]
{\includegraphics[width=.75\textwidth]{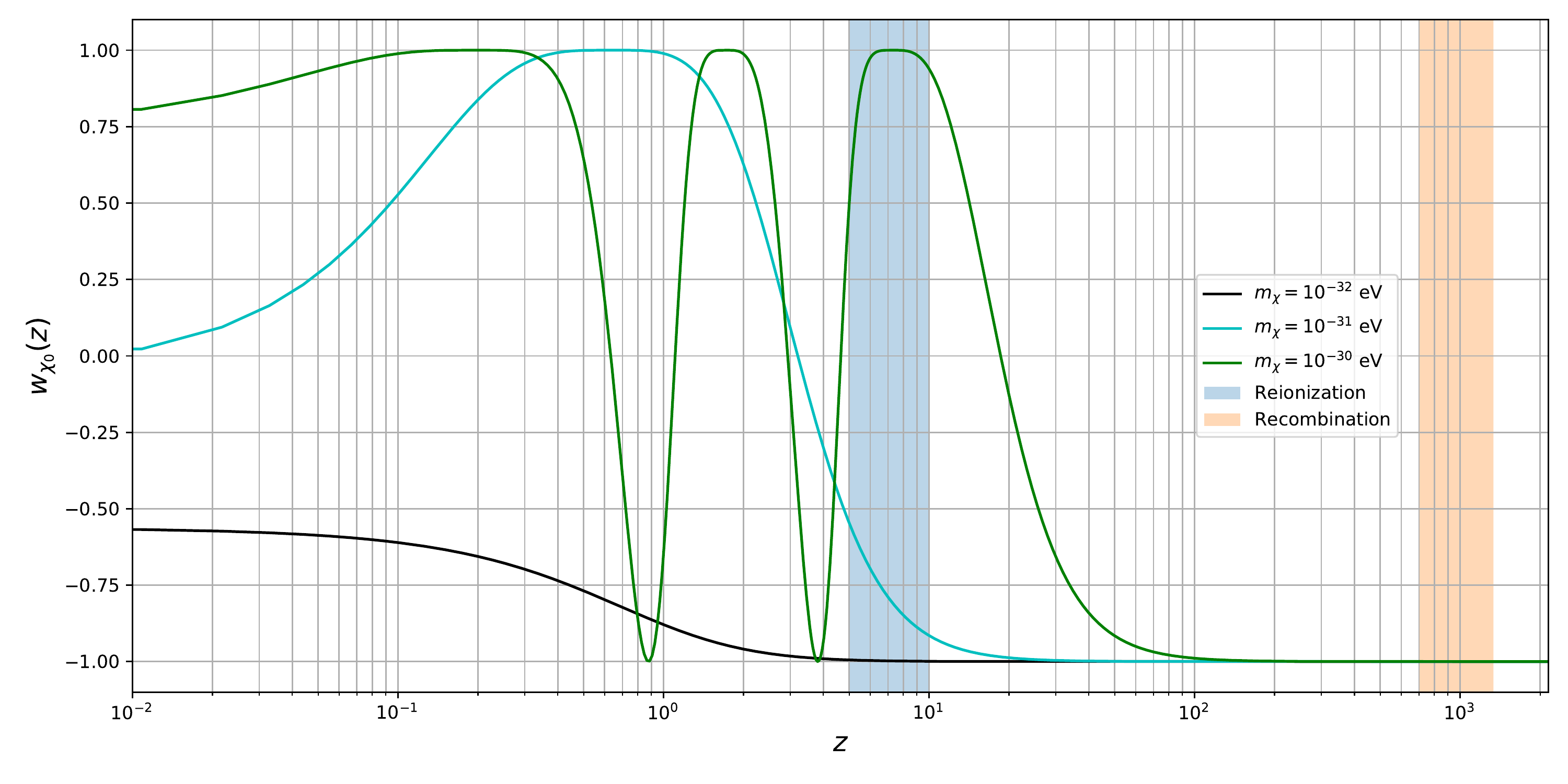}} \\
\subfloat[][\label{fig:Omega_chi}Comparison between the density parameter of the axion-like field, $\Omega_{\chi_0}\equiv(\rho_{\chi_0}a^2)/(3m_{Pl}^2\mathcal{H}^2)$, and the density parameters of the other cosmic species (radiation $\Omega_r$, matter $\Omega_m$ and cosmological constant $\Omega_{\Lambda}$).]
{\includegraphics[width=.75\textwidth]{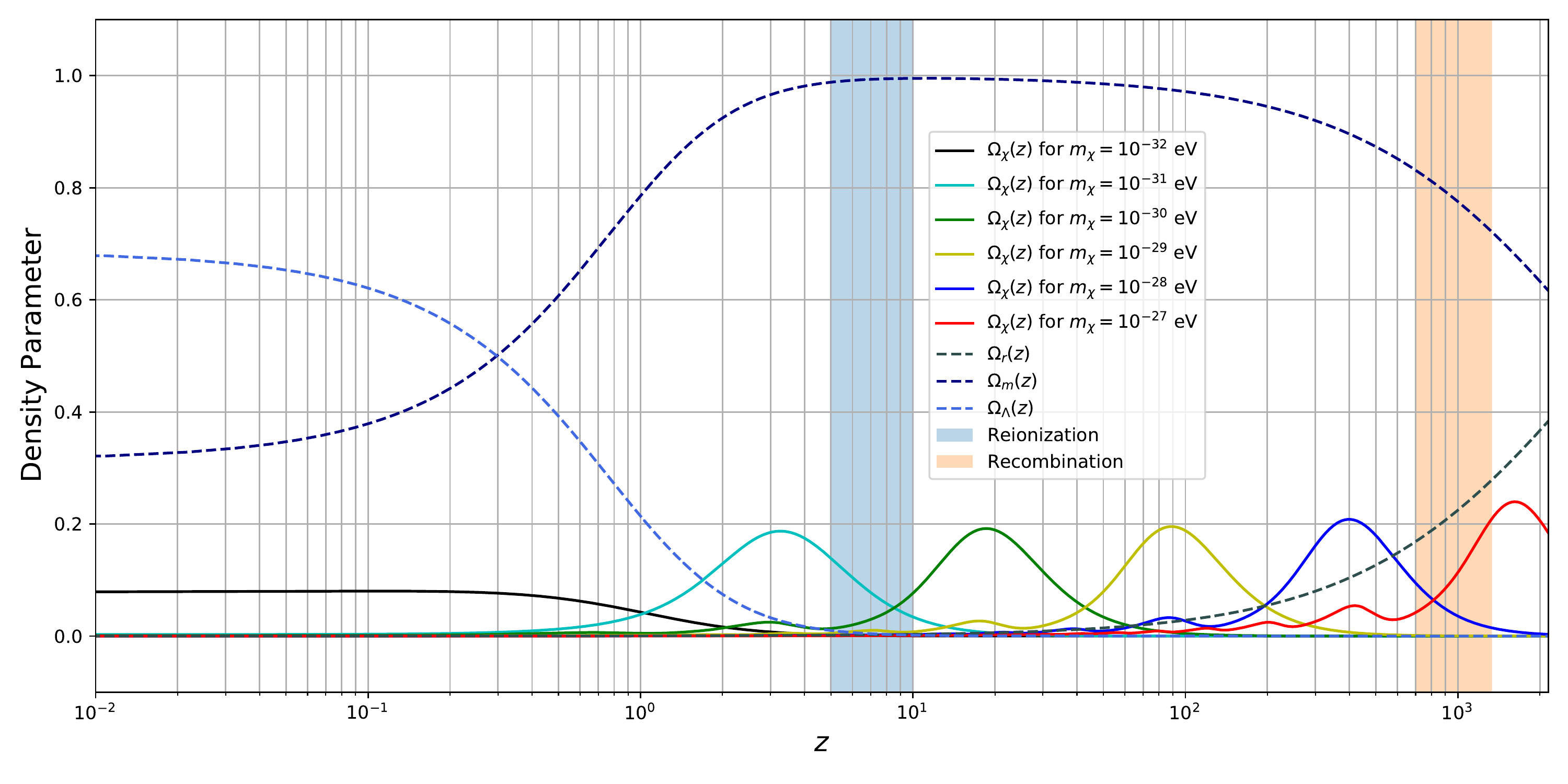}}
\caption{Background axion quantities as functions of the redshift $z$ for the model defined by Eq.~\eqref{eqn:quintessence_potential}. The numerical computation has been performed for several values of the field mass $m_{\chi}$ with $\chi_{0}^{\text{ini}}= m_{pl}$, $\chi_{0}^{\prime\,\text{ini}}=0$, and the fiducial values of the $\Lambda$CDM parameters \cite{aghanim2020planck}. The colored regions have been numerically evaluated by using the \texttt{HyRec} algorithm for recombination and the $\texttt{tanh()}$ model for reionization, respectively.}
\label{fig:axion_background}
\end{figure}

The evolution of $\chi_0$ is governed by its equation of motion (EOM), which is found by varying the kinematic part of the action in Eq.~\eqref{eqn:Chern-Simons} with respect to $\chi_0$ in the FLRW metric\footnote{Let us notice that the Chern-Simons term does not contribute to the background EOM of the axion field. This is due to the fact that
\begin{equation}
\tilde{F}^{\mu\nu}F_{\mu\nu}\propto\epsilon^{\mu\nu\rho\sigma}F_{\mu\nu}F_{\rho\sigma}\propto\epsilon^{\mu\nu\rho\sigma}\partial_{\mu}A_{\nu}\partial_{\rho}A_{\sigma}=0,
\end{equation}
since at the background level the electromagnetic field $A_{\mu}$ should be homogeneous, forcing the $4$-divergences to be time derivatives. However, the Levi-Civita symbol identically vanishes for $\mu=\rho=0$.}:
\begin{equation}
\label{eqn:background_equation}
\chi_0^{\prime\prime}+2\mathcal{H}\chi_0^{\prime}+a^2\frac{\mathrm{d}V}{\mathrm{d}\chi_0}=0,
\end{equation}
where $a$ is the scale factor of the Universe, $\mathcal{H}$ is the conformal Hubble parameter, and the prime denotes the derivative with respect to the conformal time. In order to track the evolution of the scalar field, in this paper we specialize our analysis to the case in which $\chi$ is a quintessence-like field playing the role of early dark energy with the following potential \cite{caldwell2011cross,capparelli2020cosmic,zhai2020effects,murai2022isotropic,fujita2021detection}:
\begin{equation}
\label{eqn:quintessence_potential}
V(\chi_{0})=m_{\chi}^2M_{Pl}^2\left[1-\cos\left(\frac{\chi_0}{M_{Pl}}\right)\right]^2 \, ,
\end{equation}
where $M_{Pl}$ is the Planck mass and $m_{\chi}$ is a parameter which defines the mass of the field $\chi$. We have modified the Boltzmann code \texttt{CLASS} \cite{blas2011cosmic} in order to solve Eq.~\eqref{eqn:background_equation} and obtain the plot shown in Fig.~\ref{fig:chi_background} for different values of the mass parameter $m_{\chi}$. By looking at it, we can observe that smaller the field mass is, slower its evolution is \cite{nakatsuka2022cosmic,lee2022probing}. 

It is possible to qualitatively identify five different phenomenological regimes:
\begin{itemize}
\item if $m_{\chi}\gg\SI{e-27}{\electronvolt}$, there is no isotropic cosmic birefringence, since $\chi_0$ is equal to zero both at recombination and at reionization;
\item if $\SI{e-29}{\electronvolt}\ll m_{\chi}\lesssim\SI{e-27}{\electronvolt}$, then only recombination contributes to isotropic cosmic birefringence, since $\chi_0(\tau_{\text{reco}})\ne0=\chi_0(\tau_0)$ but $\chi_0(\tau_{\text{reio}})=0=\chi_0(\tau_0)$;
\item if $\SI{e-32}{\electronvolt}\ll m_{\chi}\lesssim\SI{e-29}{\electronvolt}$, then both recombination and reionization contribute to isotropic cosmic birefringence, with different rotation angles;
\item if $m_{\chi}\lesssim\SI{e-32}{\electronvolt}$, then both recombination and reionization contribute to isotropic cosmic birefringence, but the rotation angle is the same, since $\chi_0(\tau_{\text{reco}})=\chi_0(\tau_{\text{reio}})$;
\item if $m_{\chi}\ll\SI{e-32}{\electronvolt}$, there is again no isotropic cosmic birefringence, since $\chi_0(\tau_{\text{reco}})=\chi_0(\tau_{\text{reio}})=\chi_0(\tau_0)$.
\end{itemize}
From this analysis we can infer that an appreciable birefringence effect occurs only within a finite window of masses ($m_{\chi}\in[10^{-32}\si{\electronvolt},\,10^{-27}\si{\electronvolt}]$). As we will see in the next sections, the anisotropic contribution to cosmic birefringence allows one to probe higher values for the axion mass. Note that independently on the axion mass, the field experiences a slow-roll phase at early times (and so at high redshifts): this is due to the fact that we have taken $\chi_0^{\prime,\,\text{ini}}=0$, which is a natural requirement for $\chi_0$ if we want it to behave as early dark energy. To understand this, let us compute the energy-momentum tensor for the scalar field: this can be done by varying the scalar kinematic sector of the action in Eq.~\eqref{eqn:Chern-Simons} with respect to the metric tensor
\begin{equation}
T_{\mu\nu}\equiv-\frac{2}{\sqrt{-g}}\frac{\delta\mathcal{S}_{\chi}}{\delta g^{\mu\nu}}=\partial_{\mu}\chi\partial_{\nu}\chi-g_{\mu\nu}\left[\frac{1}{2}g^{\alpha\beta}\partial_{\alpha}\chi\partial_{\beta}\chi+V(\chi)\right]\, ,
\end{equation}
so that we can easily get the (background) energy density and pressure of the scalar field as
\begin{equation}
\rho_{\chi_0}\equiv-T^0_{\ 0}=\frac{1}{2a^2}\chi_0^{\prime2}+V(\chi),\qquad\qquad\qquad\qquad P_{\chi_0}\equiv\frac{1}{3}\delta^{i}_{\ j}T^{j}_{\ i}=\frac{1}{2a^2}\chi_0^{\prime 2}-V(\chi_0).
\end{equation}
Therefore the equation of state of the homogeneous scalar field $\chi_0$ is given by
\begin{equation}
w_{\chi_0}\equiv\frac{P_{\chi_0}}{\rho_{\chi_0}}=\frac{\chi_0^{\prime 2}-2a^2V(\chi_0)}{\chi_0^{\prime 2}+2a^2V(\chi_0)},
\end{equation}
and it is equal to $w_{\chi_0}=-1$, i.e. it behaving as dark energy, when $\chi_0$ is frozen, i.e. for $\chi_0^{\prime}=0$. Hence, the requirement $w_{\chi_0}^{\text{ini}}=-1$ implies $\chi_0^{\prime,\,\text{ini}}=0$. In Fig.~\ref{fig:w_chi} we have used our modified version of \texttt{CLASS} to track the evolution of the equation of state for the scalar field $\chi$. Let us notice that this axion-like field has a rich phenomenology, since different masses imply a different nature of the field: e.g., if $m_{\chi}\simeq\SI{e-31}{\electronvolt}$, although the field may have been dark energy in the early Universe, today it may be contributing to dark matter, since $w_{\chi_0}(z=0)\simeq0$.

Therefore, a tomographic approach to isotropic cosmic birefringence can trace the axion field dynamics and how it depends on the parameters of the axion potential, such as the mass $m_{\chi}$. However, as we are going to show in the next sections, important additional and complementary information can be found by looking also at the anisotropic counterpart of cosmic birefringence, for which a tomographic approach is something new and never explored in literature to our knowledge.

However, let us just mention that in this paper, for sake of simplicity, we have not added the contribution due to the axion field in the conformal Hubble parameter $\mathcal{H}$ appearing in Eq.~\eqref{eqn:background_equation}: this is a sort of ``spectator field approximation'' for the axion field, which is valid only if $\rho_{\chi_0}$ makes $\chi_0$ a subdominant cosmic species \cite{nakatsuka2022cosmic,lee2022probing}, and the field initials conditions are chosen to satisfy this condition, as shown in Fig.~\ref{fig:Omega_chi}.

\section{\label{sec:Aniso}Axion-Induced Anisotropic Cosmic Birefringence}
If the axion-like field $\chi$ introduced in Eq.\eqref{eqn:Chern-Simons} is not homogeneous (and depends also on the spatial position) then the birefringence phenomenon is no more isotropic, encoding an anisotropic signature. In order to see this, let us write the field $\chi$ as the sum of the homogeneous background term we studied in Sec.~\ref{sec:chi_0} plus an inhomogeneous perturbation:
\begin{equation}
\label{eqn:perturbation}
\chi(\tau,\mathbf{x})=\chi_0(\tau)+\delta\chi(\tau,\mathbf{x}),
\end{equation}
where $\mathbf{x}$ is a vector of the comoving spatial coordinates. Let us now consider a photon emitted at the time $\tau$ in the point $\mathbf{x}$ and observed at the time $\tau_0$ in the point $\mathbf{x}_0$: then if we put the observer at the origin of the coordinate system and we assume a spatially flat Universe, the starting point of the photon's path is related to the coming direction of the electromagnetic wave $-\versor{n}$ via a simple relation:
\begin{equation}
\mathbf{x}=(\tau_0-\tau)\versor{n}.
\end{equation}
The most general expression for the birefringence angle reads \cite{li2008cosmological}
\begin{equation}
\int_{\text{emission}}^{\text{observation}}\mathrm{d}x^{\mu}\partial_{\mu}\alpha=\frac{\lambda}{2f}\int_{\text{emission}}^{\text{observation}}\mathrm{d}x^{\mu}\partial_{\mu}\chi\implies \int_{\text{emission}}^{\text{observation}}\mathrm{d}\alpha=\frac{\lambda}{2f}\int_{\text{emission}}^{\text{observation}}\mathrm{d}\chi,
\end{equation}
that is, by substituting Eq.~\eqref{eqn:perturbation} and integrating over the photon's null path (see e.g. \cite{zhao2014fluctuations}),
\begin{equation}
\label{eqn:aniso_angle}
\begin{split}
\alpha(\tau,\versor{n})=\frac{\lambda}{2f}\left[\chi_0(\tau_0)-\chi_0(\tau)-\delta\chi(\tau,\Delta\tau\,\versor{n})\right],
\end{split}
\end{equation}
where we have defined $\Delta\tau\equiv\tau_0-\tau$ and neglected the final value of the fluctuation $\delta\chi(\tau_0,\mathbf{0})$, since it only gives rise to ridefinition of $\chi_0(\tau_0)$. Let us notice that in absence of the field's fluctuations Eq.~\eqref{eqn:aniso_angle} just reduces to the isotropic case. It is then convenient to redefine the birefringence angle for the photons emitted at the conformal time $\tau$ in order to isolate the anisotropic contribution as
\begin{equation}
\alpha(\tau,\versor{n})=\alpha_0(\tau)+\delta\alpha(\tau,\versor{n})\, , 
\end{equation}
where $\alpha_{0}(\tau)$ is the isotropic birefringence angle defined in Eq.~\eqref{eqn:gen_iso_angle}, whereas
\begin{equation}
\label{eqn:anisotropic_birefringence}
\delta\alpha(\tau,\versor{n})\equiv-\frac{\lambda}{2f}\delta\chi(\tau,\Delta\tau\,\versor{n})
\end{equation}
is the anisotropic one. For each fixed $\tau$ it is now possible to expand the anisotropic birefringence angle over the sky through a standard spherical harmonics decomposition, see e.g. \cite{cai2022impact},
\begin{equation}
\delta\alpha(\tau,\versor{n})=\sum_{\ell m}\alpha_{\ell m}(\tau)Y_{\ell m}(\versor{n}),
\end{equation} 
since $\delta\alpha$ is a scalar quantity. It follows that the harmonic coefficients $\alpha_{\ell m}$ are defined at any emission time as
\begin{equation}
\label{eqn:anisotropies}
\alpha_{\ell m}(\tau)=-\frac{\lambda}{2f}\int\mathrm{d}^2\hat{n}\,Y_{\ell m}^{*}(\versor{n})\,\delta\chi(\tau,\Delta\tau\,\versor{n}).
\end{equation}
Now we move to the Fourier space,
\begin{equation}
\delta\chi(\tau,\Delta\tau\versor{n})=\int\frac{\mathrm{d}^3k}{(2\pi)^3}\,e^{i\Delta\tau\mathbf{k}\cdot\versor{n}}\delta\chi(\tau,\mathbf{k}),
\end{equation}
and we adopt the plane wave-expansion, 
\begin{equation}
e^{i\Delta\tau\mathbf{k}\cdot\versor{n}}=4\pi\sum_{LM}i^{L}j_{L}(k\Delta\tau)Y_{LM}^*(\versor{k})Y_{LM}(\versor{n}),
\end{equation}
with $j_{\ell}(k\Delta\tau)$ being the $\ell$-th spherical Bessel function, so that we can rewrite Eq.~\eqref{eqn:anisotropies} as
\begin{equation}
\label{eqn:harmonic_birefringence}
\alpha_{\ell m}(\tau)=-\frac{4\pi\lambda i^{\ell}}{2f}\int\frac{\mathrm{d}^3k}{(2\pi)^3}\,Y_{\ell m}^*(\versor{k})j_{\ell}(k\Delta\tau)\delta\chi(\tau,\mathbf{k}),
\end{equation}
where we have exploited the orthonormality property of the spin-weighted spherical harmonics:
\begin{equation}
\label{eqn:orthonormality}
\int\mathrm{d}^2\hat{n}\,_{s}Y_{\ell m}^*(\versor{n})\,_{s}Y_{LM}(\versor{n})=\delta_{\ell L}\delta_{mM}\quad\forall s\in\mathbb{Z}.
\end{equation}
Armed with the expression given in Eq.~\eqref{eqn:harmonic_birefringence}, we can then compute the angular power spectrum of anisotropic cosmic birefringence:
\begin{equation}
\begin{split}
\langle\alpha_{\ell_1 m_1}^*(\tau_1)&\,\alpha_{\ell_2 m_2}(\tau_2)\rangle=\\
&=\left(-\frac{4\pi\lambda}{2f}\right)^2i^{\ell_2-\ell_1}\int\frac{\mathrm{d}^3k_1\,\mathrm{d}^3k_2}{(2\pi)^6}\,\,Y_{\ell_1 m_1}(\versor{k}_1)Y_{\ell_2m_2}^*(\versor{k}_2)j_{\ell_1}(k_1\Delta\tau_1)j_{\ell_2}(k_1\Delta\tau_2)\expval{\delta\chi^*(\tau_1,\mathbf{k}_1)\delta\chi(\tau_2,\mathbf{k}_1)},
\end{split}
\end{equation}
where $\expval{\dots}$ stands for ensemble average.

\subsection{Dynamics of the Axion Fluctuations}
In order to further proceed we need a suitable expression for the power spectrum of the field fluctuations $\delta\chi$: here we assume a statistically isotropic model of inflation and adiabatic initial conditions\footnote{See e.g. \cite{lee2006constraints,lee2019dark} for a discussion about isocurvature modes as initial conditions instead.}. In such a way, it is possible to define the two-point correlation function for the axion field fluctuations as
\begin{equation}
\expval{\delta\chi^*(\tau_1,\mathbf{k}_1)\delta\chi(\tau_2,\mathbf{k}_2)}=\frac{2\pi^2}{k_1^3}\mathcal{P}_{\mathcal{R}}(k_1)T_{\delta\chi}(\tau_1,k_1)T_{\delta\chi}(\tau_2,k_2)(2\pi)^3\delta^{(3)}(\mathbf{k}_1-\mathbf{k}_2),
\end{equation} 
where, since anisotropic cosmic birefringence is sourced just by scalar perturbations, $\mathcal{P}_{\mathcal{R}}(k)$ is the dimensionless power spectrum of the comoving curvature perturbation\footnote{The fact that the field fluctuation $\delta\chi$ is related to the comoving curvature perturbation $\mathcal{R}$ (for adiabatic conditions) will be clarified very soon, after deriving the equation of motion for the axion perturbation.} $\mathcal{R}$ (equal to scalar amplitude $A_s$ in the scale-invariant case), whereas $T_{\delta\chi}(\tau,k)$ is a proper transfer function for the field fluctuation $\delta\chi$ with the role of evolving the perturbation from early primordial epoch to the given time $\tau$. Therefore, we obtain \cite{li2008cosmological,capparelli2020cosmic,caldwell2011cross,zhao2014fluctuations}
\begin{equation}
\label{eqn:alpha-alpha}
\expval{\alpha_{\ell_1 m_1}^*(\tau_1)\,\alpha_{\ell_2 m_2}(\tau_2)}=4\pi\left(\frac{\lambda}{2f}\right)^2\int\frac{\mathrm{d}k_1}{k_1}\,\mathcal{P}_{\mathcal{R}}(k_1)j_{\ell_1}(k_1\Delta\tau_1)j_{\ell_2}(k_1\Delta\tau_2)T_{\delta\chi}(\tau_1,k_1)T_{\delta\chi}(\tau_2,k_1)\delta_{\ell_1\ell_2}\delta_{m_1m_2}.
\end{equation}
In analogy with the standard CMB observables, we can parameterize the amplitude of the angular power-spectrum in Eq.~\eqref{eqn:alpha-alpha} as
\begin{equation}
\expval{\alpha_{\ell_1 m_1}^*(\tau_x)\,\alpha_{\ell_2 m_2}(\tau_z)}=C_{\ell_1}^{\alpha\alpha}\big|_{xz}\delta_{\ell_1\ell_2}\delta_{m_1m_2},
\end{equation}
where now $x$ and $z$ are labels for the different epochs\footnote{In our case they refer to recombination or reionization.}. The angular power spectrum of anisotropic birefringence is then given as
\begin{equation}
\label{eqn:Caa}
C_{\ell}^{\alpha\alpha}\big|_{xz}=4\pi\int\frac{\mathrm{d}k}{k}\,\mathcal{P}_{\mathcal{R}}(k)\Delta_{\alpha,\ell}(k,\tau_x)\Delta_{\alpha,\ell}(k,\tau_z),
\end{equation} 
where we have defined a time-integrated transfer function\footnote{For sake of simplicity, here we write a Dirac delta by doing a sort of abuse of notation, just to stress that the anisotropic birefringence angle is proportional to the inhomogeneous field fluctuation evaluated at the time of photon emission. However, our code does not evaluate $\delta\chi$ only at $\tau_{\text{reco}}$ or $\tau_{\text{reio}}$ but in practice convolves it with the photon visibility function $g(\tau)$ in the neighborhood of the recombination and reionization peak, so that we can isolate the recombination contribution from the one from reionization epoch. This approach is not only numerically more accurate but also more physically correct since of course, we have an important amount of photons emitted also in a finite range of $\tau$ close to the peaks.}:
\begin{equation}
\label{eqn:Delta_alpha}
\Delta_{\alpha,\ell}(k,\tau)\equiv-\frac{\lambda}{2f}\int^{\tau_0}_{0}\mathrm{d}\tilde{\tau}\,\delta(\tilde{\tau}-\tau)T_{\delta\chi}(\tilde{\tau},k)j_{\ell}[k(\tau_0-\tilde{\tau})]\, .
\end{equation}
Moreover, it is possible to consider also cross-correlations between cosmic birefringence with CMB temperature $T$ and $E$ polarization modes \cite{caldwell2011cross,capparelli2020cosmic,zhai2020effects,greco2022cosmic}:
\begin{align}
\expval{\alpha_{\ell_1 m_1}^*(\tau_x)\,a_{T,\ell_2 m_2}(\tau_z)}&=C_{\ell_1}^{\alpha T}\big|_{xz}\delta_{\ell_1\ell_2}\delta_{m_1m_2}\, , \\
\expval{\alpha_{\ell_1 m_1}^*(\tau_x)\,a_{E,\ell_2 m_2}(\tau_z)}&=C_{\ell_1}^{\alpha E}\big|_{xz}\delta_{\ell_1\ell_2}\delta_{m_1m_2}\, , 
\end{align}
where, in analogy with Eq.~\eqref{eqn:Caa}, the cross-spectra are given as
\begin{align}
\label{eqn:Cat}
C_{\ell}^{\alpha T}\big|_{xz}=4\pi\int\frac{\mathrm{d}k}{k}\,\mathcal{P}_{\mathcal{R}}(k)\Delta_{\alpha,\ell}(k,\tau_x)\Delta_{T,\ell}(k,\tau_z)\, , \\
\label{eqn:Cae}
C_{\ell}^{\alpha E}\big|_{xz}=4\pi\int\frac{\mathrm{d}k}{k}\,\mathcal{P}_{\mathcal{R}}(k)\Delta_{\alpha,\ell}(k,\tau_x)\Delta_{E,\ell}(k,\tau_z)\, , 
\end{align}
whereas the $C_{\ell}^{\alpha B}$ cross-correlation is predicted to be zero, since the $B$ modes of CMB polarization are instead sourced just by tensor perturbations (see however \cite{workinprogress}). 

As can be seen by looking at Eq.~\eqref{eqn:anisotropic_birefringence}, the physical source of anisotropic cosmic birefringence is the field fluctuation $\delta\chi$: thus, in order to completely determine the transfer function $T_{\delta\chi}$, it is necessary to solve the equation of motion for $\delta\chi$, which is found by varying the scalar sector of the action in Eq.~\eqref{eqn:Chern-Simons} with respect to $\delta\chi$. By adopting the conformal Newtonian gauge for the linear scalar perturbations \cite{ma1995cosmological},
\begin{equation}
g_{\mu\nu}=a^2(\tau)\begin{pmatrix}
	-[1+2\Psi(\tau,\mathbf{x})] & 0 \\
	0 & [1-2\Phi(\tau,\mathbf{x})]\delta_{ij}
\end{pmatrix}
\end{equation}
and by working in Fourier space, one finds that at linear order in perturbation theory the equation of motion for the axion field fluctuations reads
\begin{equation}
\label{eqn:delta_chi}
\begin{split}
\delta\chi^{\prime\prime}+2\mathcal{H}\delta\chi^{\prime}+\left(k^2+a^2\frac{\mathrm{d}^2V}{\mathrm{d}\chi_0^2}\right)\delta\chi&=(\chi_0^{\prime\prime}+2\mathcal{H}\chi_0^{\prime})(3\Phi+\Psi)+\chi_0^{\prime}(3\Phi^{\prime}+\Psi^{\prime})+\\
&\quad\qquad+a^2\frac{\mathrm{d}V}{\mathrm{d}\chi_0}(3\Phi-\Psi)+\frac{i\lambda}{f}\epsilon^{ijk}A_{0,i}^{\prime}k_j\delta A_k,
\end{split}
\end{equation}
where we have decomposed the electromagnetic field in analogy with Eq.~\eqref{eqn:perturbation} as
\begin{equation}
A_{\mu}(\tau,\mathbf{x})=A_{0,\mu}(\tau)+\delta A_{\mu}(\tau,\mathbf{x}).
\end{equation}
Moreover, we can expand the spatial part of the electromagnetic field perturbation over a basis of the two polarization states, i.e.
\begin{equation}
\delta A_i(\tau,\mathbf{k})=\sum_{s=\pm1}\varepsilon_i^{(s)}(\versor{k})\,\delta A_{(s)}(\tau,\mathbf{k}),
\end{equation}
where $\varepsilon_i^{(s)}(\versor{k})$ is a divergenceless polarization vector in order to satisfy the Lorenz gauge condition, $\delta^{ij}\hat{k}_i\varepsilon_j^{(s)}(\versor{k})=0$, and that has some other important properties \cite{shiraishi2011cmb}:
\begin{equation}
\label{eqn:pol_vec}
\left[\varepsilon_i^{(s)}(\versor{k})\right]^*=\varepsilon_i^{(-s)}(\versor{k})=\varepsilon_i^{(s)}(-\versor{k}),\quad\qquad\delta^{ij}\varepsilon_i^{(s_1)}\varepsilon_j^{(-s_2)}(\versor{k})=\delta_{s_1s_2},\quad\qquad\epsilon^{ijk}\hat{k}_j\varepsilon_k^{(s)}(\versor{k})=-is\,\delta^{ij}\varepsilon_{j}^{(s)}(\versor{k}).
\end{equation}
By substituting Eq.~\eqref{eqn:background_equation} in Eq.~\eqref{eqn:delta_chi} and by exploiting the properties collected in Eq.~\eqref{eqn:pol_vec}, we obtain
\begin{equation}
\label{eqn:equation_of_motion}
\delta\chi^{\prime\prime}+2\mathcal{H}\delta\chi^{\prime}+\left(k^2+a^2\frac{\mathrm{d}^2V}{\mathrm{d}\chi_0^2}\right)\delta\chi=\chi_{0}^{\prime}(3\Phi^{\prime}+\Psi^{\prime})-2a^2\frac{\mathrm{d}V}{\mathrm{d}\chi_0}\Psi+\frac{\lambda}{f}\delta^{ij}A_{0,i}^{\prime}\sum_{s=\pm1}s\,\varepsilon_j^{(s)}\delta A_{(s)},
\end{equation}
where the $A_{0,i}^{\prime}$ is the derivative with respect to the conformal time of the $i$-th component of the background term of the electromagnetic field. However, in order to preserve statistical isotropy, the electromagnetic field must have a vanishing vacuum expectation value, i.e. $\mathbf{A}_{0}=0$, and for this reason from now on we are going to disregard the vector contribution to the axion equation of motion. The form of Eq.~\eqref{eqn:equation_of_motion} clarifies our previous statement about the fact that anisotropic cosmic birefringence is sourced just by scalar perturbations, since $\Phi$ and $\Psi$ appearing on the r.h.s. of Eq.~\eqref{eqn:equation_of_motion} are the two scalar gauge-invariant Bardeen gravitational potentials. Indeed, this explains how the field fluctuation $\delta\chi$ is related to the metric perturbations, and so, more generally, to the gauge-invariant comoving curvature perturbation $\mathcal{R}$ (for adiabatic perturbations). 

\subsection{Angular Power Spectra of Anisotropic Cosmic Birefringence}
\begin{figure}
\centering
\subfloat[][\label{fig:alpha-alpha}Auto-Correlation of ACB. The shaded regions are excluded by the present constraints from \textit{Planck} PR3 \cite{bortolami2022planck} with  the \texttt{Commander} component separation method (whose analysis refers to multipoles in the $2\le\ell\le24$ range), from ACTPol \cite{namikawa2020atacama} (whose analysis refers to multipoles in the $20\le\ell\le2048$ range) and from SPTpol \cite{bianchini2020searching} (whose analysis refers to multipoles in the $50\le\ell\le2000$ range), respectively.]
{\includegraphics[width=.65\textwidth]{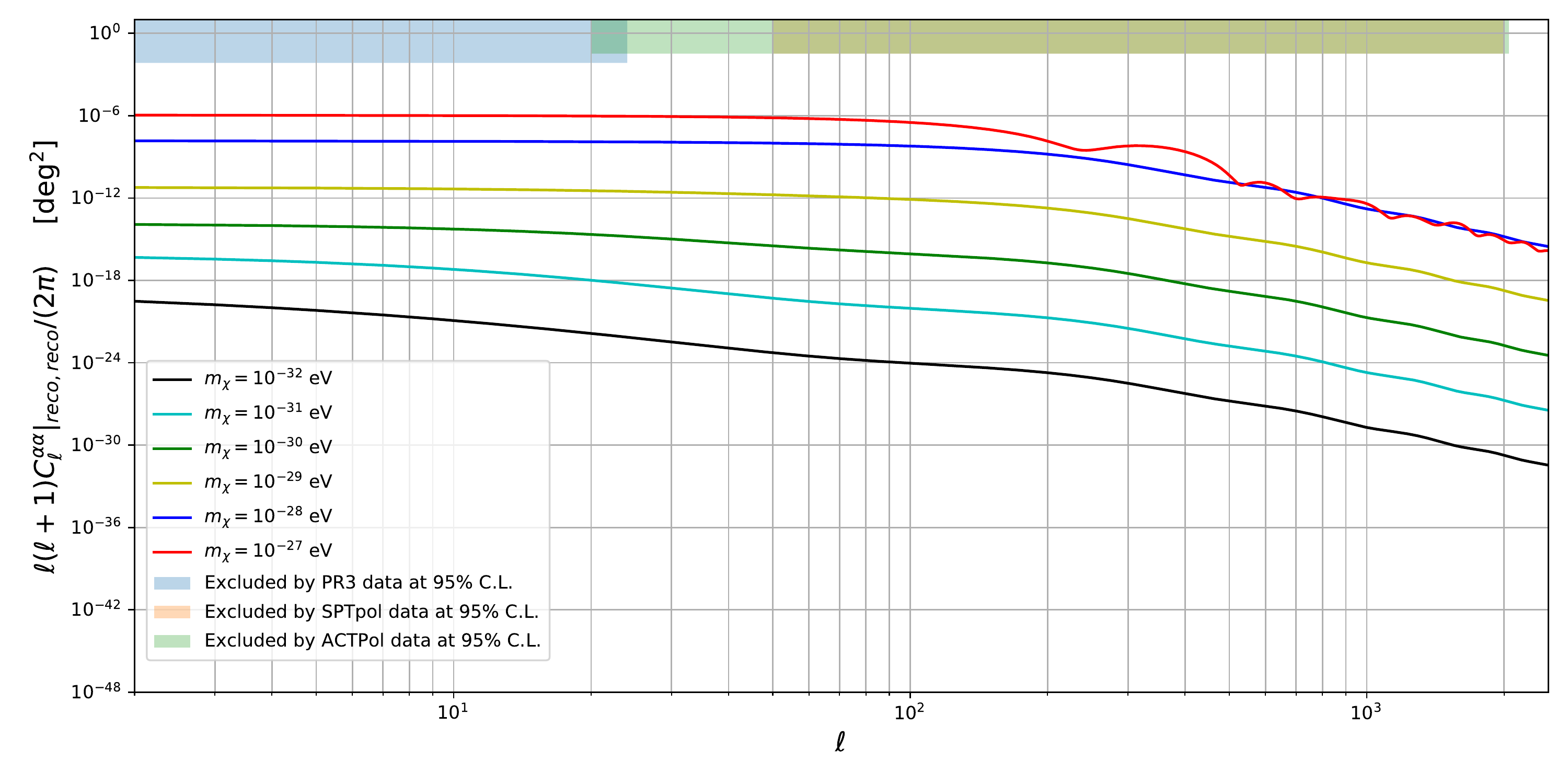}} \\
\subfloat[][\label{fig:alpha-T}Absolute value of the cross-correlation of ACB with CMB temperature. The shaded regions are excluded by the present constraints from \textit{Planck} PR3 \cite{bortolami2022planck}  with the \texttt{Commander} component separation method (whose analysis refers to multipoles in the $2\le\ell\le24$ range), and from SPTpol \cite{bianchini2020searching} (whose analysis refers to multipoles in the $100\le\ell\le2000$ range), respectively.]
{\includegraphics[width=.65\textwidth]{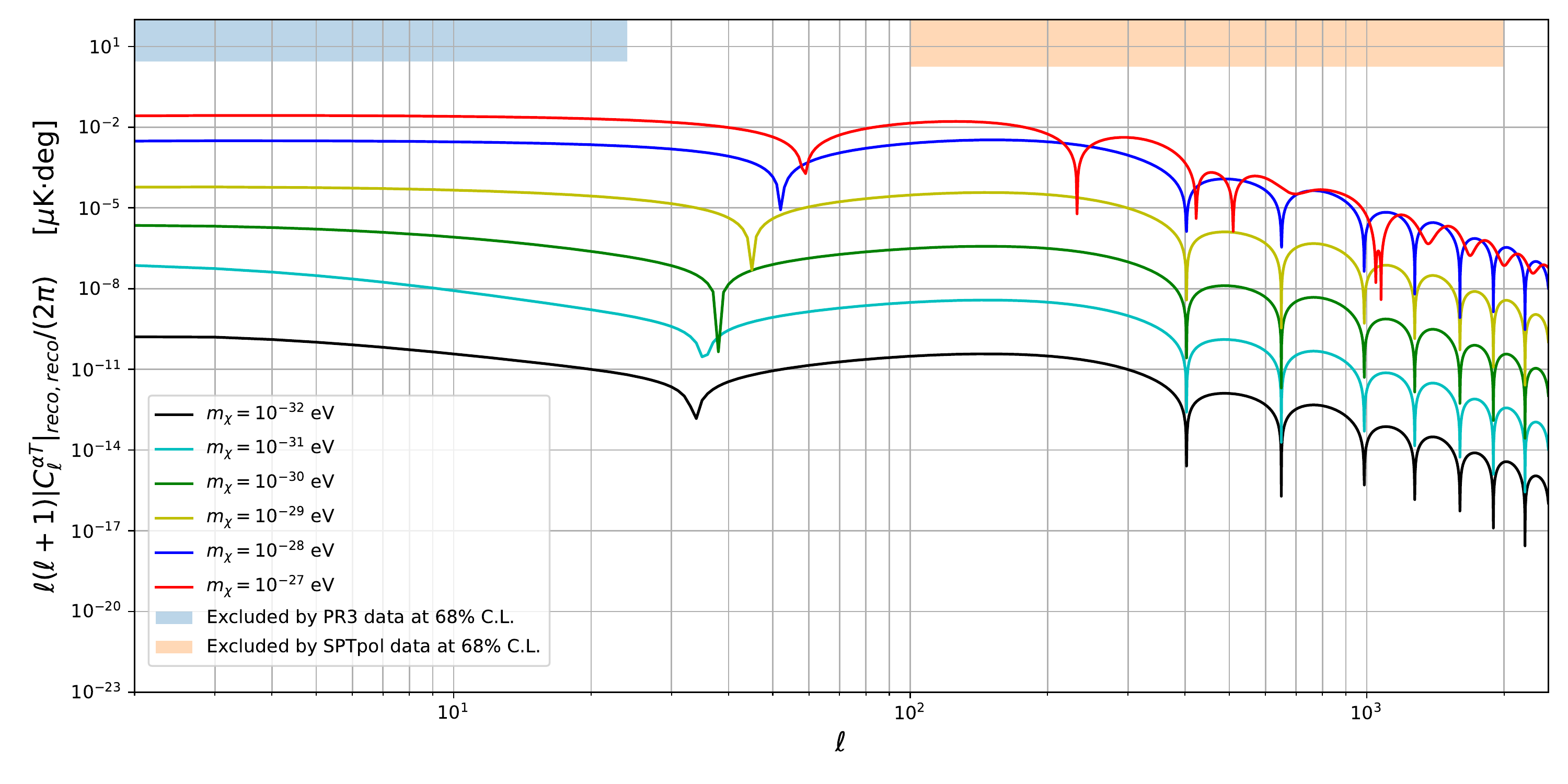}} \\
\subfloat[][\label{fig:alpha-E}Absolute value of the cross-correlation of ACB with $E$ modes of CMB polarization. The shaded region is that excluded by the present constraints from \textit{Planck} PR3 \cite{bortolami2022planck} with  the \texttt{Commander} component separation method (whose analysis refers to multipoles in the $2\le\ell\le23$ range).]
{\includegraphics[width=.65\textwidth]{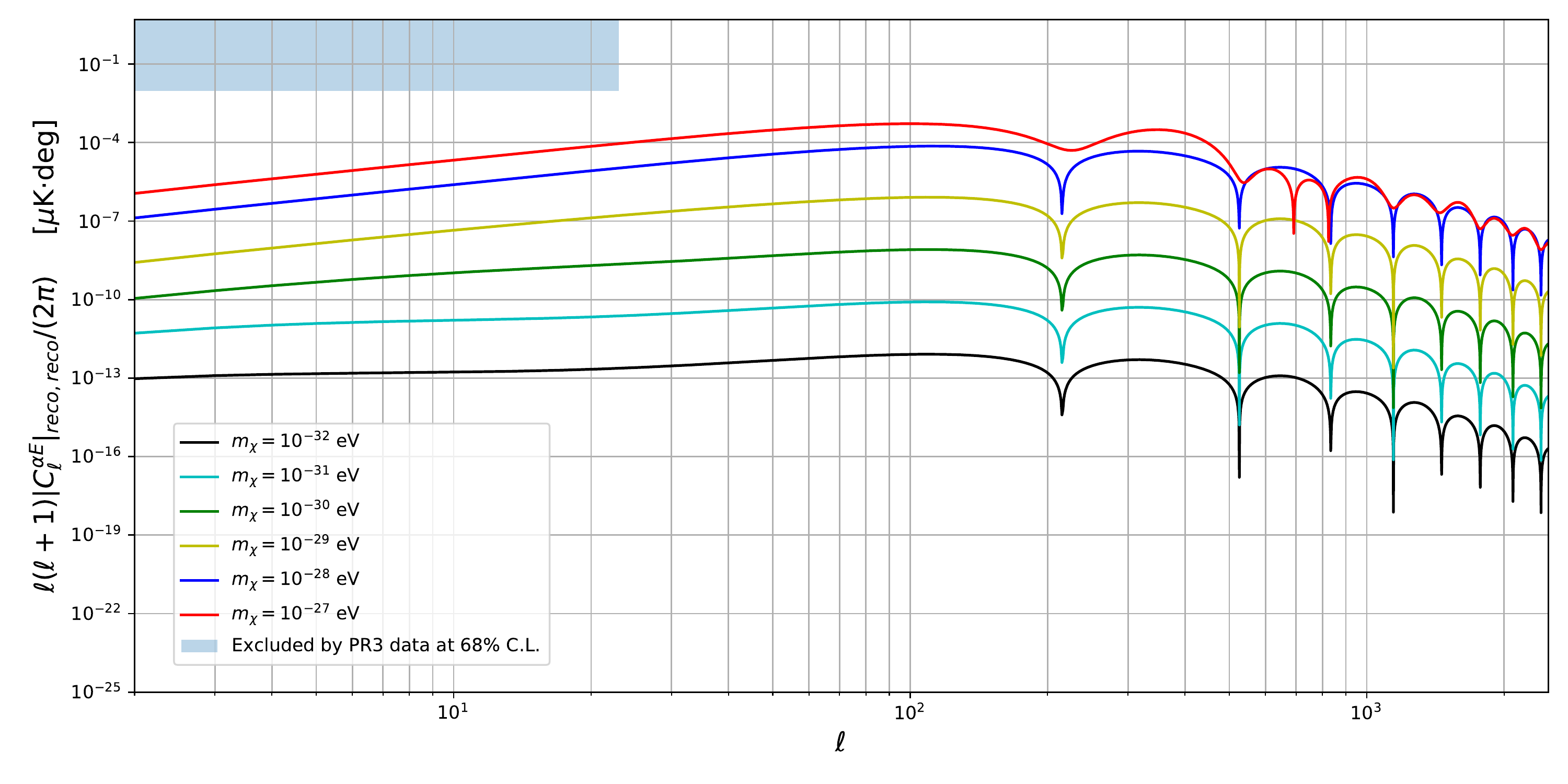}}
\caption{Angular power spectra involving anisotropic cosmic birefringence for the model defined by Eq.~\eqref{eqn:quintessence_potential}. The numerical computation has been performed for several values of the field mass, by taking $\lambda/f=\SI{e-18}{\giga\electronvolt^{-1}}$, $\chi_{0}^{\text{ini}}= m_{Pl}$ (reduced Planck mass), $\chi_{0}^{\prime\,\text{ini}}=0$, and for the fiducial values of the $\Lambda$CDM parameters provided in \cite{aghanim2020planck}.}
\label{fig:aniso_spectra}
\end{figure}

We have modified \texttt{CLASS} \cite{blas2011cosmic}, in order to implement anisotropic cosmic birefringence: our code evaluates the angular power spectra defined in Eq.~\eqref{eqn:Caa} and Eqs.~\eqref{eqn:Cat}-\eqref{eqn:Cae} by solving\footnote{To be precise, our code solves Eq.~\eqref{eqn:equation_of_motion} in the synchronous gauge, 
\begin{equation}
g_{\mu\nu}=a^2(\tau)\begin{pmatrix}
-1 & 0 \\
			0 & [\delta_{ij}+h_{ij}(\tau,\mathbf{x})]
		\end{pmatrix},
	\end{equation}
	where
	\begin{equation}
		h_{ij}(\tau,\mathbf{x})=\frac{1}{3}h(\tau,\mathbf{x})\delta_{ij}+\left[\partial_i\partial_j-\frac{1}{3}\delta_{ij}\nabla^2\right]\mu(\tau,\mathbf{x})+\left[\partial_i\mathcal{A}_j(\tau,\mathbf{x})+\partial_j\mathcal{A}_i(\tau,\mathbf{x})\right]+h_{ij}^T(\tau,\mathbf{x}),
	\end{equation}
with $h$, $\mu$, $\mathcal{A}_i$, $h_{ij}^T$ being the scalar, divergenceless vector and divergenceless transverse tensor perturbations, respectively. In this gauge Eq.~\eqref{eqn:equation_of_motion} reads
\begin{equation}
\delta\chi^{\prime\prime}+2\mathcal{H}\delta\chi^{\prime}+\left(k^2+a^2\frac{\mathrm{d}^2V}{\mathrm{d}\chi_0^2}\right)\delta\chi=-\frac{1}{2}\chi_0^{\prime}h^{\prime}+\frac{\lambda}{f}\delta^{ij}A_{0,i}^{\prime}\sum_{s=\pm1}s\,\varepsilon_k^{(s)}\delta A_{(s)},
\end{equation}
After solving the equation above, our code moves to the Newtonian conformal gauge via a gauge transformation, $\delta\chi^{\text{New}}=\delta\chi^{\text{Syn}}+\alpha^{\prime}\chi_{0}^{\prime}$, where $\alpha$ is not the birefringence angle but just the time-component of the four-vector inducing the gauge-transformation \cite{ma1995cosmological,caldwell2011cross}, which is in turn computed by \texttt{CLASS}. This is done because the time-derivative of the potential $\Psi$ appearing in Eq.~\eqref{eqn:equation_of_motion} is not implemented in \texttt{CLASS}.} the perturbed EOM we derived in Eq.~\eqref{eqn:equation_of_motion} together with the Einstein equations for the two scalar potentials $\Psi$ and $\Phi$. $C^{\alpha\alpha}_{\ell}$, $C_{\ell}^{\alpha T}$ and $C_{\ell}^{\alpha E}$ given in Eq.\eqref{eqn:Caa} and Eqs.~\eqref{eqn:Cat}-\eqref{eqn:Cae} are plotted in Fig.~\ref{fig:aniso_spectra} for $\tau_1=\tau_2=\tau_{\text{reco}}$, where for now we have not considered the signal coming from reionization in order to be consistent with the current observational constraints.

As anticipated in Sec.~\ref{sec:Intro}, by looking at Fig.~\ref{fig:aniso_spectra}, we can remark that for the potential defined in Eq.~\eqref{eqn:quintessence_potential}, larger the scalar field mass is, larger the spectra's amplitudes are\footnote{We have checked that using a quadratic potential, $V(\chi)=m^2_{\chi}\chi^2/2$, considered e.g. in \cite{nakatsuka2022cosmic,lee2022probing}, does not qualitatively affect our plots and the main conclusions do not change in a significant way.}: this is a peculiar behavior of anisotropic cosmic birefringence. Note that, as we discussed in Sec.~\ref{sec:chi_0}, an heavy axion field would implies no isotropic birefringence, since the background field $\chi_0$ starts to oscillate before the recombination time; on the contrary Fig.~\ref{fig:aniso_spectra} is telling us an interesting aspect: the more massive the axion field, the more amplitude is enhanced, allowing us to investigate a wider range of masses.

There is in fact a clear physical explanation for this intriguing phenomenon. Indeed, since anisotropic birefringence is sourced by the perturbations of the axion field, the fact that the larger the axion mass is, larger the field fluctuations are, can seem counter-intuitive, since we expect a heavy field to fluctuate less than a light one. In order to clarify this aspect, we now adopt a semi-analytical approach. Let us focus on Eq.~\eqref{eqn:harmonic_birefringence}: the birefringence angle is related to the value of the axion fluctuation at the recombination or at the reionization time, and both of these two events occur during the matter-dominated epoch (as can be seen by also looking at Fig.~\ref{fig:Omega_chi}). Hence, we just need to solve Eq.~\eqref{eqn:equation_of_motion} in that epoch, when the two Newtonian potentials can be taken as approximately constant in time , so that $\Psi^{\prime}\sim\Phi^{\prime}\sim0$. Therefore, Eq.~\eqref{eqn:equation_of_motion} just reduces to
	\begin{equation}
		\label{eqn:delta_chi_matter_dominated}
		\delta\chi^{\prime\prime}+2\mathcal{H}\delta\chi^{\prime}+\left(k^2+a^2\frac{\mathrm{d}^2V}{\mathrm{d}\chi_0^2}\right)\delta\chi=-2a^2\frac{\mathrm{d}V}{\mathrm{d}\chi_0}\Psi.
\end{equation}

Now we restrict our analysis to the case in which $m_{\chi}$ is small but non-vanishing. Indeed, let us consider $m_{\chi}\lesssim\SI{e-31}{\electronvolt}$ so that, as can seen by looking at Fig.~\ref{fig:chi_background}, we can safely adopt the slow-roll approximation. It can be proved that in this regime, if one assumes adiabatic initial conditions (as we did) the axion field perturbation that solves Eq.~\eqref{eqn:delta_chi_matter_dominated} can be written as \cite{li2008cosmological,caldwell2011cross,capparelli2020cosmic,zhai2020effects}
	\begin{equation}
		\label{eqn:fluctuation}
		\delta\chi(\tau,k)\propto a^2\tau^2\frac{\mathrm{d}V}{\mathrm{d}\chi_0}\Psi(\tau,k).
	\end{equation}
	Thanks to the presence of a non-vanishing potential, it becomes clear why the scalar perturbations of the metric are able to source the cross-correlation between anisotropic cosmic birefringence and the other scalar-sourced CMB observables. Moreover, we can see that the strength of this coupling is proportional to $\mathrm{d}V/\mathrm{d}\chi_0$, which is in turn proportional to $m_{\chi}^2$. Therefore, this explains why increasing the axion mass implies an enhancement of the spectra's amplitude for anisotropic cosmic birefringence.

Furthermore, let us see what happens if instead we set $m_{\chi}=0$ in Eq.~\eqref{eqn:quintessence_potential}: it follows that the axion potential becomes exactly equal to zero, so that Eq.~\eqref{eqn:delta_chi_matter_dominated} further simplifies to
	\begin{equation}
		\label{eqn:vacuum_solution}
		\delta\chi^{\prime\prime}+2\mathcal{H}\delta\chi^{\prime}+a^2k^2\delta\chi=0.
	\end{equation}
	Hence, a vanishing potential in the matter-dominated epoch prevents the axion-like field perturbations from having any correlation with perturbations in the matter/radiation density \cite{caldwell2011cross}, and this is evident because of the absence of any source term in the right-hand side of Eq.~\eqref{eqn:vacuum_solution}. A trivial consequence of this is that, if the axion mass is exactly zero, we would have $C_{\ell}^{\alpha T}=C_{\ell}^{\alpha E}=0$.

The theoretical results shown in Fig.~\ref{fig:aniso_spectra} are consistent with those derived in \cite{capparelli2020cosmic,zhai2020effects}, and are compared with the most recent measurements, in particular with the analysis of \textit{Planck} PR3 data in~\cite{bortolami2022planck}, which gives the observational constraints on the scale-invariant angular correlations of anisotropic birefringence using the \texttt{Commander} component separation method. Other important constraints on anisotropic cosmic birefringence come by former analysis of the \textit{Planck} mission \cite{contreras2017constraints,gruppuso2020planck}, and by other experiments, such as  ACTPol \cite{namikawa2020atacama}, SPTpol \cite{bianchini2020searching}, Bicep-Keck \cite{ade2017bicep2,keck2022line}, Polarbear \cite{ade2015polarbear} and WMAP \cite{gluscevic2012first}.

Although a full comparison of theory with observations is beyond the purpose of this paper, we nevertheless just point out that a joint investigation of all the cross-spectra of cosmic birefringence can allow us to extract fundamental information\footnote{Although in this paper we are going to focus on the two-point correlation functions, let us just mention that it has been shown recently that also the three-point angular correlation functions of anisotropic cosmic birefringence can be seen as really promising cosmological observables \cite{greco2022cosmic}.} about the field mass $m_{\chi}$ or in general on the field potential $V(\chi)$. 

For example, already from these results from anisotropic birefringence, one can conclude that, within the context of these models and for the fixed value of the axion-photon coupling parameter $\lambda/f=\SI{e-18}{\giga\electronvolt^{-1}}$ that we chose in this paper, spectra with higher masses than those we considered in this paper are generally excluded by the current observational constraints, as shown in Fig.~\ref{fig:aniso_spectra}. A direct implication of that is that for sure a non-vanishing isotropic birefringence should be produced, since that the range of masses that are excluded by the anisotropic signal forces the background axion field to evolve in time as shown in Fig.~\ref{fig:chi_background}. Further investigations about the free parameters of the underlying axion model are left for future works \cite{greco_prep}.

\section{\label{sec:Tomo}Tomographic Analysis of ACB}
As we mentioned in Sec.~\ref{sec:Intro}, CMB photons were mainly emitted at two different epochs: recombination and reionization, and so we can define the harmonic coefficients for the two anisotropic birefringence angles as
\begin{equation}
\alpha_{\ell m}^{\text{reco}}\equiv\alpha_{\ell m}(\tau_{\text{reco}})\, ,\qquad\qquad\qquad\qquad\alpha_{\ell m}^{\text{reio}}\equiv\alpha_{\ell m}(\tau_{\text{reio}}).
\end{equation}
According to \cite{sherwin2021cosmic,nakatsuka2022cosmic}, we can generalize Eq.~\eqref{eqn:rotation} in such a way that the Stokes parameters measured after the rotation induced by cosmic birefringence are given by
\begin{equation}
\left[Q(\versor{n})\pm iU(\versor{n})\right]_{\text{rot}}=\sum_{x=\text{reio},\text{reco}}\left[Q_x(\versor{n})\pm iU_x(\versor{n})\right]e^{\pm2i\left[\alpha_{0}(\tau_x)+\delta\alpha(\tau_x,\versor{n})\right]}\, . 
\end{equation}

\subsection{Impact on CMB Anisotropies}
We can now focus on understanding how cosmic birefringence from the two epochs (recombination and reionization) affects the CMB power spectra: by definition, since the linear combination of Stokes parameters $(Q\pm iU)$ behaves as a spin-2 field, it can be projected on the celestial sky via a proper set of spin-weighted spherical harmonics,
\begin{equation}
\left[Q(\versor{n})\pm iU(\versor{n})\right]_{\text{rot}}=\sum_{\ell m}a^{\text{rot}}_{\pm2,\ell m}\,_{\pm2}Y_{\ell m}(\versor{n}),
\end{equation}
so that the related harmonic coefficients are then defined as \cite{greco2022cosmic}
\begin{equation}
a^{\text{rot}}_{\pm2,\ell m}=\sum_{x}\sum_{LM}\int\mathrm{d}^2\hat{n}\,\,_{\pm2}Y_{\ell m}^*(\versor{n})\,_{\pm2}Y_{L M}(\versor{n})\,a^x_{\pm2,LM}e^{\pm2i\left[\alpha_{0}(\tau_x)+\delta\alpha(\tau_x,\versor{n})\right]},
\end{equation}
where $a^x_{\pm2,LM}$ are the unrotated harmonic coefficients for the Stokes parameters, i.e. those that we would have in absence of cosmic birefringence:
\begin{equation}
a^{x}_{\pm2,\ell m}=\int\mathrm{d}^2\hat{n}\,_{\pm2}Y_{\ell m}^*(\versor{n})\,\left[Q_x(\versor{n})\pm iU_x(\versor{n})\right].
\end{equation}
The harmonic coefficients of the $E$ and $B$ polarization modes are given as \cite{hu1997cmb}
\begin{equation}
a_{\pm 2,\ell m}\equiv-(a_{E,\ell m}\pm ia_{B,\ell m}),
\end{equation}
and so, after some trivial calculations, it is easy to show that the following formula holds true \cite{greco2022cosmic}:
\begin{equation}
\begin{pmatrix}
a_{E,\ell m}^{\text{rot}} \\
a_{B,\ell m}^{\text{rot}}
\end{pmatrix}=\sum_{x}\begin{pmatrix}
a_{E,\ell m}^{\text{rot},x} \\
a_{B,\ell m}^{\text{rot},x}
\end{pmatrix}=
\sum_{x}\sum_{s=\pm2}\frac{e^{is\alpha_{0}(\tau_x)}}{2}\sum_{LM}\int\mathrm{d}\hat{n}\,_sY_{\ell m}^*(\versor{n})\,_sY_{LM}(\versor{n})\begin{pmatrix}
1 & is/2 \\
-is/2 & 1
\end{pmatrix}
\begin{pmatrix}
a_{E,LM}^{x} \\
a_{B,LM}^{x}
\end{pmatrix}e^{is\delta\alpha(\tau_x,\versor{n})},
\end{equation}
where we have adopted a short-hand notation for
\begin{equation}
a^x_{\pm 2,\ell m}\equiv-(a_{E,\ell m}^x\pm ia_{B,\ell m}^x).
\end{equation}
The harmonic coefficients of CMB temperature anisotropies instead are not affected by cosmic birefringence:
\begin{equation}
T(\versor{n})=\sum_{\ell m}a_{T,\ell m}Y_{\ell m}(\versor{n})
\end{equation}
with
\begin{equation} 
a_{T,\ell m}=\int\mathrm{d}^2\hat{n}\,Y_{\ell m}^*(\versor{n})T(\versor{n})=\sum_{x=\text{reco,reio}}a_{T,\ell m}^x.
\end{equation}
The CMB power spectra measured after rotation are then given as
\begin{align}
\label{eqn:EE}
C_{\ell,\text{rot}}^{EE}&=C_{\ell,\text{rot}}^{EE}\big|_{\text{reco,reco}}+2C_{\ell,\text{rot}}^{EE}\big|_{\text{reco,reio}}+C_{\ell,\text{rot}}^{EE}\big|_{\text{reio,reio}},\\
\label{eqn:BB}
C_{\ell,\text{rot}}^{BB}&=C_{\ell,\text{rot}}^{BB}\big|_{\text{reco,reco}}+2C_{\ell,\text{rot}}^{BB}\big|_{\text{reco,reio}}+C_{\ell,\text{rot}}^{BB}\big|_{\text{reio,reio}},\\
\label{eqn:EB}
C_{\ell,\text{rot}}^{EB}&=C_{\ell,\text{rot}}^{EB}\big|_{\text{reco,reco}}+C_{\ell,\text{rot}}^{EB}\big|_{\text{reco,reio}}+C_{\ell,\text{rot}}^{EB}\big|_{\text{reio,reco}}+C_{\ell,\text{rot}}^{EB}\big|_{\text{reio,reio}},\\
\label{eqn:TE}
C_{\ell,\text{rot}}^{TE}&=C_{\ell,\text{rot}}^{TE}\big|_{\text{reco,reco}}+C_{\ell,\text{rot}}^{TE}\big|_{\text{reco,reio}}+C_{\ell,\text{rot}}^{TE}\big|_{\text{reio,reco}}+C_{\ell,\text{rot}}^{TE}\big|_{\text{reio,reio}},\\
\label{eqn:TB}
C_{\ell,\text{rot}}^{TB}&=C_{\ell,\text{rot}}^{TB}\big|_{\text{reco,reco}}+C_{\ell,\text{rot}}^{TB}\big|_{\text{reco,reio}}+C_{\ell,\text{rot}}^{TB}\big|_{\text{reio,reco}}+C_{\ell,\text{rot}}^{TB}\big|_{\text{reio,reio}},
\end{align} 
similarly to what is done in \cite{greco2022cosmic}.

A full analytical derivation of all the components of Eqs.~\eqref{eqn:EE}-\eqref{eqn:TB} can be found in App.~\ref{App:CMB}, and we report here the results collected in Eqs.~\eqref{eqn:APP_EE}-\eqref{eqn:APP_EB} and~\eqref{eqn:APP_TE}-\eqref{eqn:APP_TB}:

\begin{equation}
\label{eqn:C_EE}
\begin{split}
C_{\ell,\text{rot}}^{EE}\big|_{xz}=&\left(1-2V_{\alpha}\big|_{xx}-2V_{\alpha}\big|_{zz}\right)\left[C^{EE}_{\ell}\big|_{xz}\cos(2\alpha_{0,x})\cos(2\alpha_{0,z})+C_{\ell}^{BB}\big|_{xz}\sin(2\alpha_{0,x})\sin(2\alpha_{0,z})\right]\\
&+\frac{2}{2\ell+1}\sum_{L_1L_2}I^{2,-2,0}_{\ell L_1L_2}\Big(C^{\alpha\alpha}_{L_2}\big|_{xz}I^{2,-2,0}_{\ell L_1L_2}\Big\{C^{EE}_{L_1}\big|_{xz}\left[\cos(2\alpha_{0,x}-2\alpha_{0,z})-(-1)^{\ell+L_1+L_2}\cos(2\alpha_{0,x}+2\alpha_{0,z})\right]\\
&\hspace{140pt}+C^{BB}_{L_1}\big|_{xz}\left[\cos(2\alpha_{0,x}-2\alpha_{0,z})+(-1)^{\ell+ L_1+L_2}\cos(2\alpha_{0,x}+2\alpha_{0,z})\right]\Big\}\\
&\hspace{80pt}+C^{\alpha E}_{L_1}\big|_{xz}C^{\alpha E}_{L_2}\big|_{zx}I^{2,0,-2}_{\ell L_1L_2}\left[\cos(2\alpha_{0,x}-2\alpha_{0,z})-(-1)^{\ell+L_1+L_2}\cos(2\alpha_{0,x}+2\alpha_{0,z})\right]\Big),
\end{split}
\end{equation}
\begin{equation}
\label{eqn:C_BB}
\begin{split}
C_{\ell,\text{rot}}^{BB}\big|_{xz}=&\left(1-2V_{\alpha}\big|_{xx}-2V_{\alpha}\big|_{zz}\right)\left[C^{BB}_{\ell}\big|_{xz}\cos(2\alpha_{0,x})\cos(2\alpha_{0,z})+C_{\ell}^{EE}\big|_{xz}\sin(2\alpha_{0,x})\sin(2\alpha_{0,z})\right]\\
&+\frac{2}{2\ell+1}\sum_{L_1L_2}I^{2,-2,0}_{\ell L_1L_2}\Big(C^{\alpha\alpha}_{L_2}\big|_{xz}I^{2,-2,0}_{\ell L_1L_2}\Big\{C^{EE}_{L_1}\big|_{xz}\left[\cos(2\alpha_{0,x}-2\alpha_{0,z})+(-1)^{\ell+L_1+L_2}\cos(2\alpha_{0,x}+2\alpha_{0,z})\right]\\
&\hspace{140pt}+C^{BB}_{L_1}\big|_{xz}\left[\cos(2\alpha_{0,x}-2\alpha_{0,z})-(-1)^{\ell+ L_1+L_2}\cos(2\alpha_{0,x}+2\alpha_{0,z})\right]\Big\}\\
&\hspace{80pt}+C^{\alpha E}_{L_1}\big|_{xz}C^{\alpha E}_{L_2}\big|_{zx}I^{2,0,-2}_{\ell L_1L_2}\left[\cos(2\alpha_{0,x}-2\alpha_{0,z})+(-1)^{\ell+L_1+L_2}\cos(2\alpha_{0,x}+2\alpha_{0,z})\right]\Big),
\end{split}
\end{equation}
\begin{equation}
\label{eqn:C_EB}
\begin{split}
C_{\ell,\text{rot}}^{EB}\big|_{xz}=&\left(1-2V_{\alpha}\big|_{xx}-2V_{\alpha}\big|_{zz}\right)\left[C^{EE}_{\ell}\big|_{xz}\cos(2\alpha_{0,x})\sin(2\alpha_{0,z})-C_{\ell}^{BB}\big|_{xz}\sin(2\alpha_{0,x})\cos(2\alpha_{0,z})\right]\\
&+\frac{2}{2\ell+1}\sum_{L_1L_2}I^{2,-2,0}_{\ell L_1L_2}\Big(C^{\alpha\alpha}_{L_2}\big|_{xz}I^{2,-2,0}_{\ell L_1L_2}\Big\{C^{BB}_{L_1}\big|_{xz}\left[\sin(2\alpha_{0,x}-2\alpha_{0,z})-(-1)^{\ell+L_1+L_2}\sin(2\alpha_{0,x}+2\alpha_{0,z})\right]\\
&\hspace{140pt}-C^{EE}_{L_1}\big|_{xz}\left[\sin(2\alpha_{0,x}-2\alpha_{0,z})+(-1)^{\ell+ L_1+L_2}\sin(2\alpha_{0,x}+2\alpha_{0,z})\right]\Big\}\\
&\hspace{80pt}-C^{\alpha E}_{L_1}\big|_{xz}C^{\alpha E}_{L_2}\big|_{zx}I^{2,0,-2}_{\ell L_1L_2}\left[\sin(2\alpha_{0,x}-2\alpha_{0,z})-(-1)^{\ell+L_1+L_2}\sin(2\alpha_{0,x}+2\alpha_{0,z})\right]\Big),
\end{split}
\end{equation}
\begin{equation}
\label{eqn:C_TE}
C^{TE}_{\ell,\text{rot}}\big|_{xz}=\left(1-2V_{\alpha}\big|_{zz}\right)\cos(2\alpha_{0,z})C^{TE}_{\ell}\big|_{xz},\hspace{290pt}
\end{equation}
\begin{equation}
\label{eqn:C_TB}
C^{TB}_{\ell,\text{rot}}\big|_{xz}=\left(1-2V_{\alpha}\big|_{zz}\right)\sin(2\alpha_{0,z})C^{TB}_{\ell}\big|_{xz},\hspace{290pt}
\end{equation}
where we have introduced the short-hand notation $\alpha_{0}(\tau_x)\equiv\alpha_{0,x}$. 

\begin{figure}
\centering
\subfloat[][\label{fig:E-E}Auto-correlation of $E$ modes of CMB polarization. The standard $\Lambda$CDM prediction for the unlensed $EE$ spectrum with no birefringence is also plotted (black dashed line).]
{\includegraphics[width=.7\textwidth]{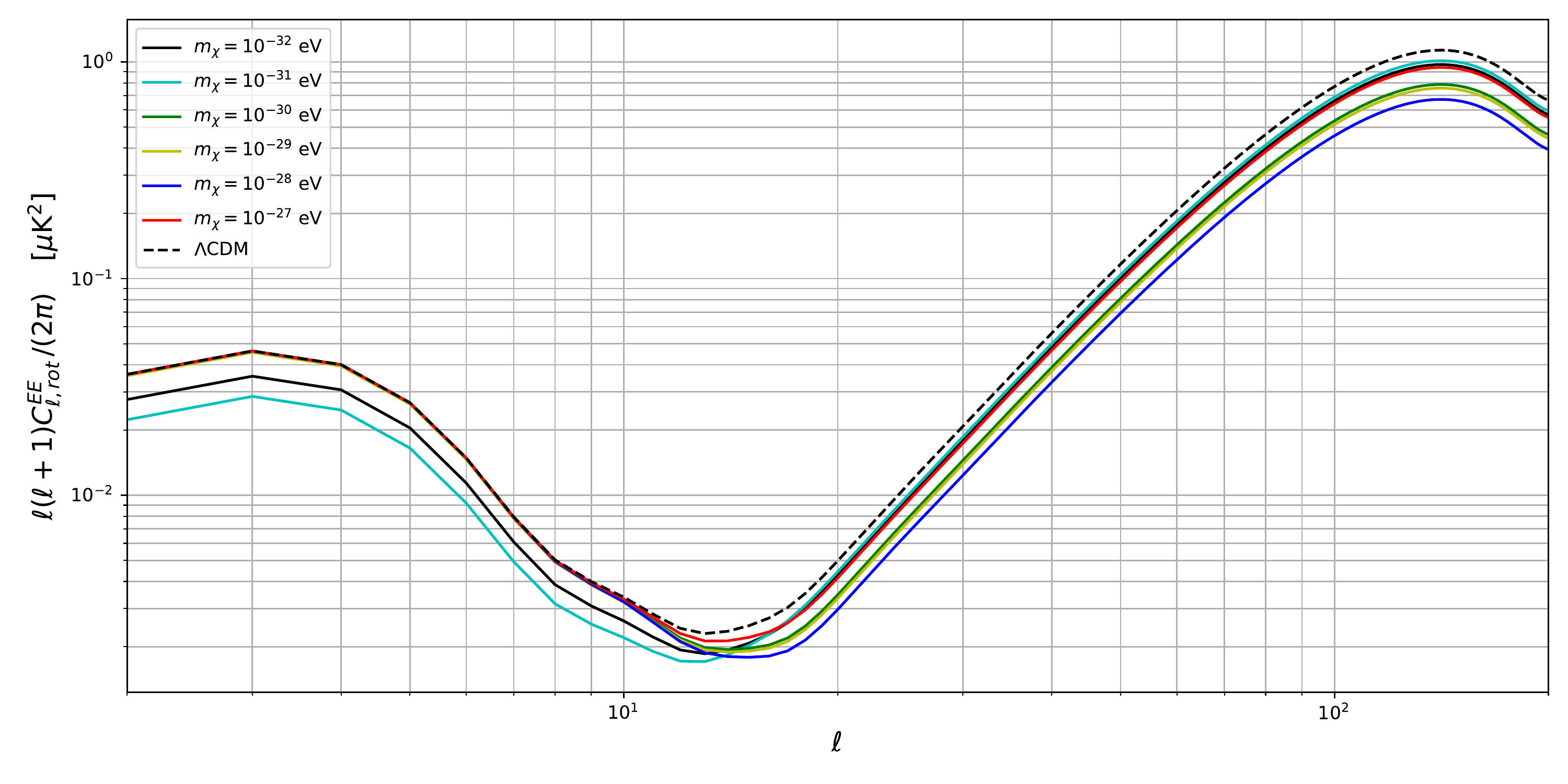}} \\
\subfloat[][\label{fig:B-B}Auto-correlation of $B$ modes of CMB polarization. In absence of primordial tensor perturbations the unlensed $BB$ spectrum with no birefringence is predicted to be zero by the standard $\Lambda$CDM model.]
{\includegraphics[width=.7\textwidth]{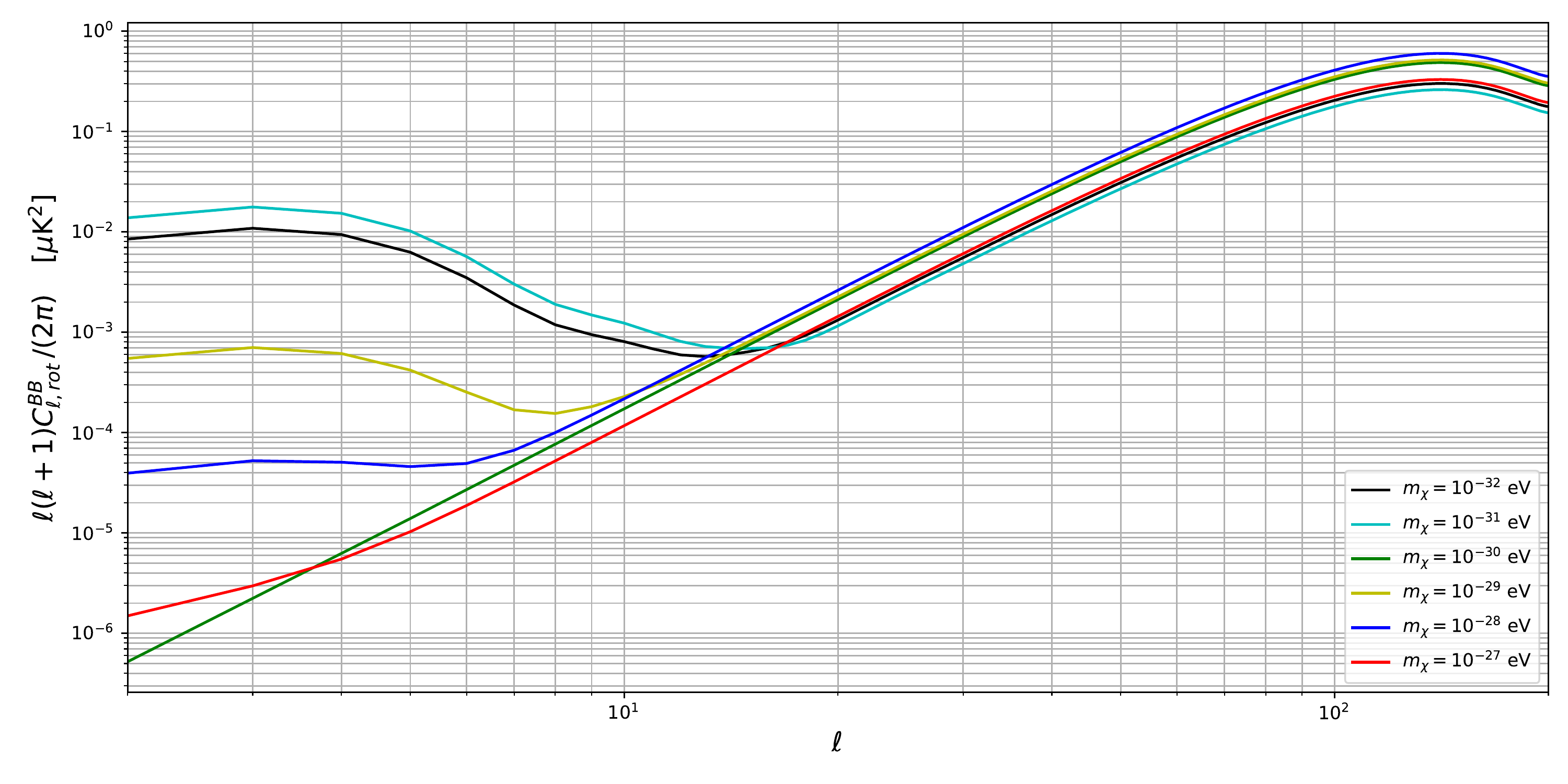}} \\
\subfloat[][\label{fig:E-B}Absolute value of the cross-correlation of $E$ and $B$ modes of CMB polarization. In absence of primordial tensor perturbations or of parity-violating mechanisms, the unlensed $EB$ spectrum with no birefringence is predicted to be zero by the standard $\Lambda$CDM model.]
{\includegraphics[width=.7\textwidth]{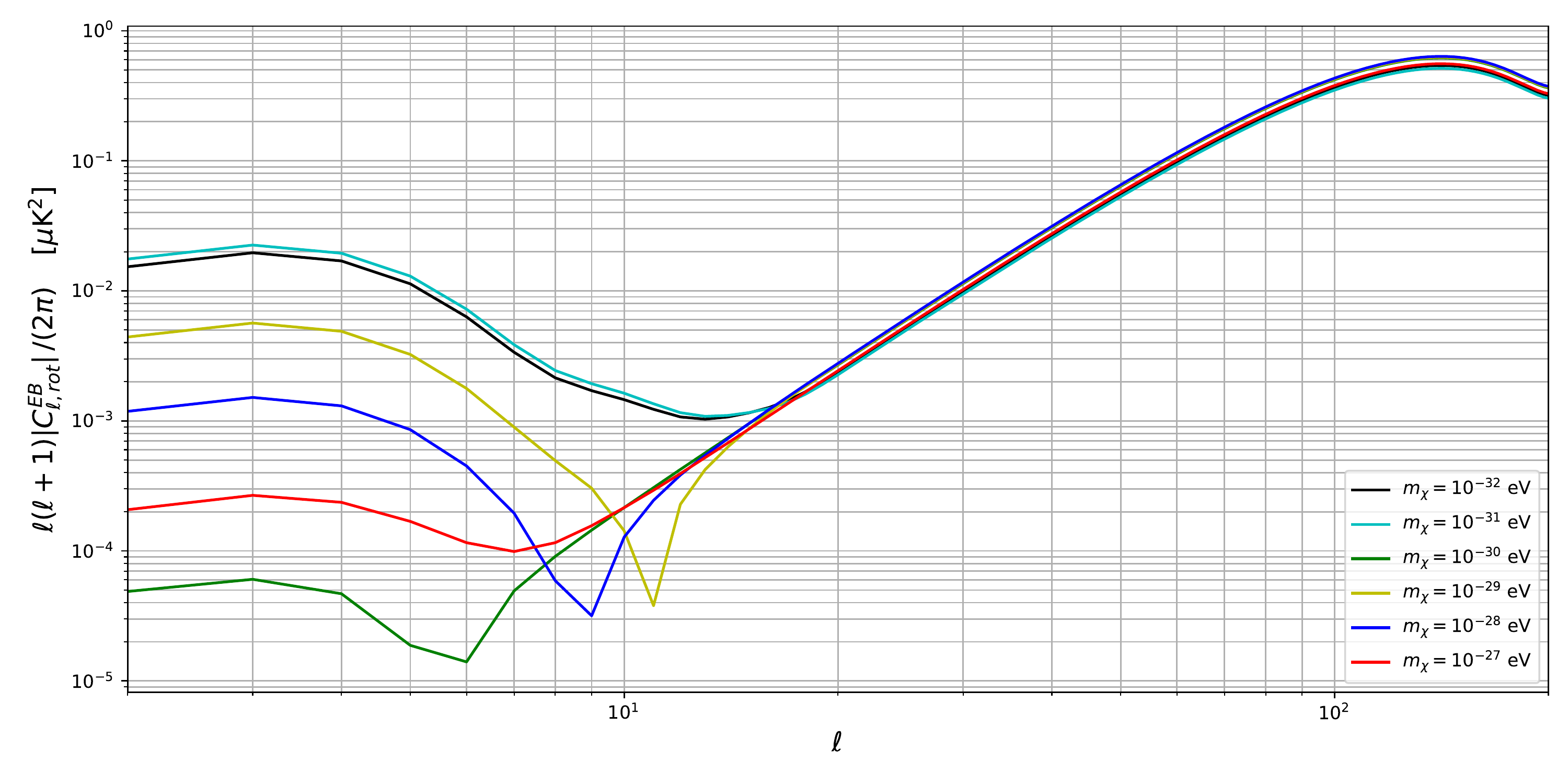}}
\caption{Unlensed angular power spectra of CMB polarization affected by isotropic and anisotropic cosmic birefringence, from recombination and reionization, for the model defined by Eq.~\eqref{eqn:quintessence_potential}. The numerical computation has been performed for several values of the field mass (straight lines), by taking $\lambda/f=\SI{e-18}{\giga\electronvolt^{-1}}$, $\chi_{0}^{\text{ini}}= m_{Pl}$ (reduced Planck mass), $\chi_{0}^{\prime\,\text{ini}}=0$, and for the fiducial values of the $\Lambda$CDM parameters provided in \cite{aghanim2020planck}. The tensor-to-scalar ratio is set equal to zero ($r=0$).}
\label{fig:pol_rot}
\end{figure}

\begin{figure}
\centering
\subfloat[][\label{fig:T-E}Absolute value of the $TE$ power spectrum. The standard $\Lambda$CDM prediction for the unlensed $TE$ spectrum with no birefringence is also plotted (black dashed line).]
{\includegraphics[width=.8\textwidth]{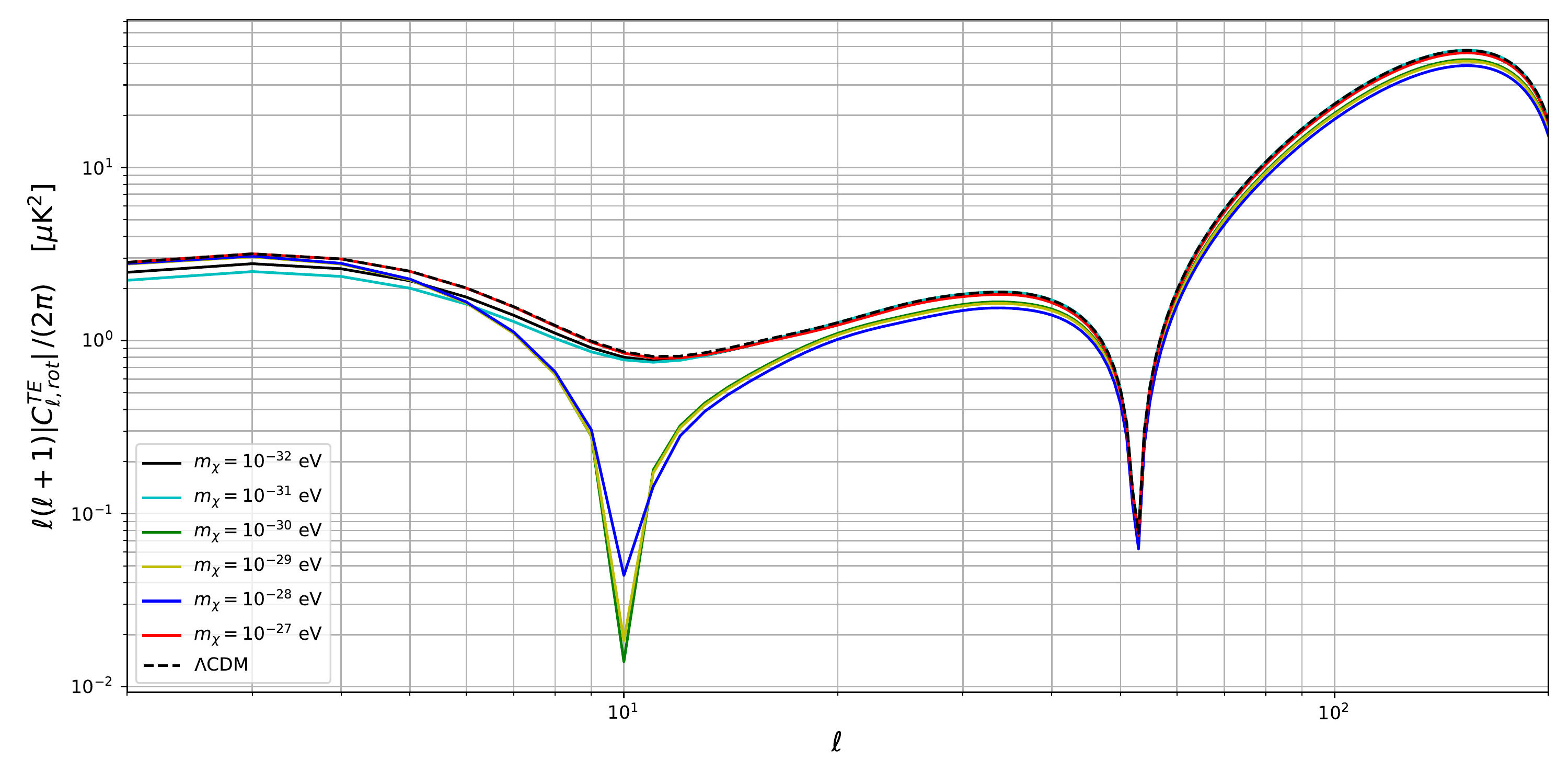}} \\
\subfloat[][\label{fig:T-B}Absolute value of the $TB$ power spectrum. In absence of primordial tensor perturbations or parity-violating mechanisms the unlensed $TB$ spectrum with no birefringence is predicted to be zero by the standard $\Lambda$CDM model.]
{\includegraphics[width=.8\textwidth]{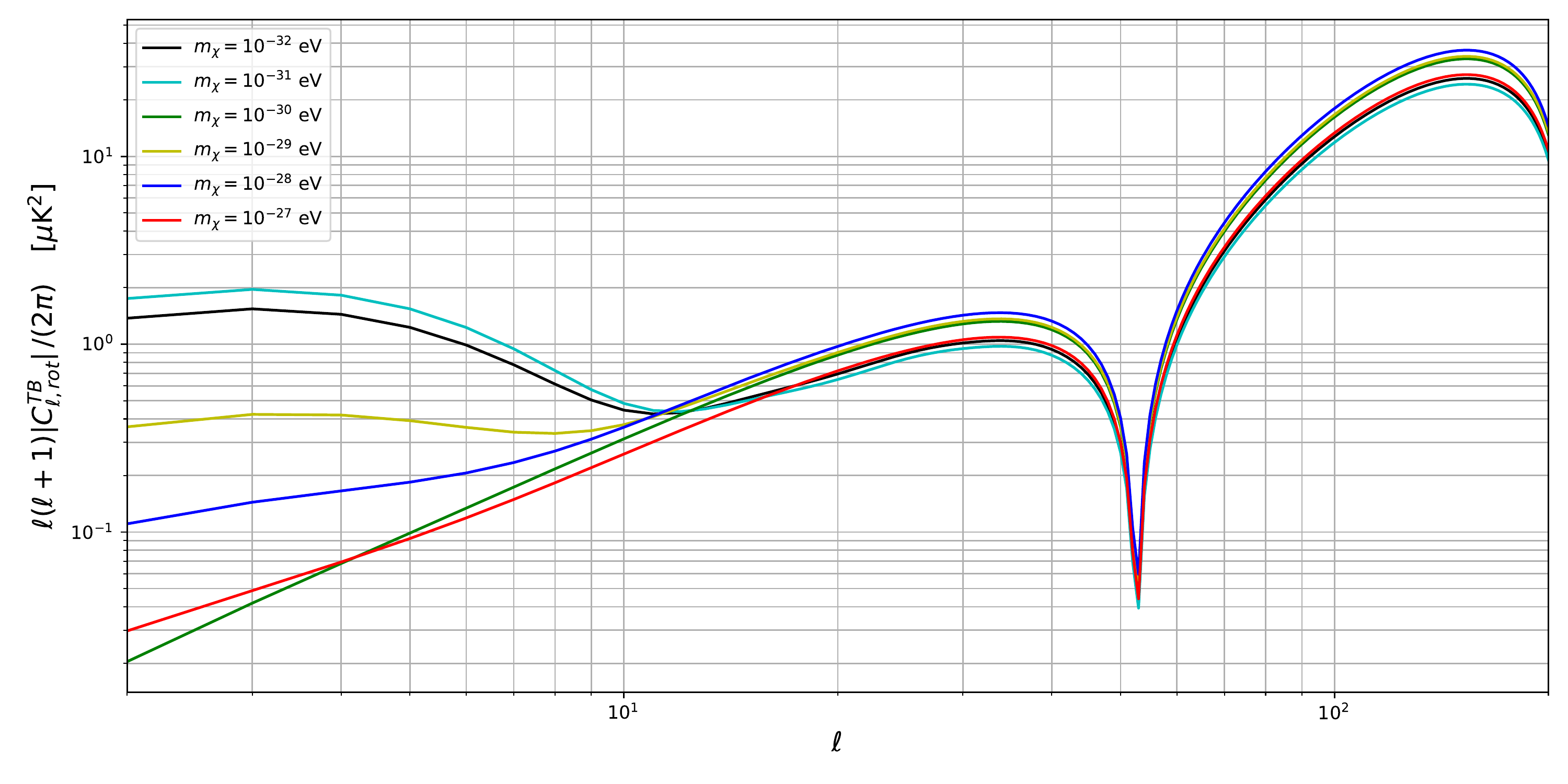}}
\caption{Unlensed angular cross-correlations of CMB temperature with $E$ and $B$ polarization modes, affected by isotropic and anisotropic cosmic birefringence, from recombination and reionization, for the model defined by Eq.~\eqref{eqn:quintessence_potential}. The numerical computation has been performed for several values of the field mass (straight lines), by taking $\lambda/f=\SI{e-18}{\giga\electronvolt^{-1}}$, $\chi_{0}^{\text{ini}}= m_{Pl}$ (reduced Planck mass), $\chi_{0}^{\prime\,\text{ini}}=0$, and for the fiducial values of the $\Lambda$CDM parameters provided in \cite{aghanim2020planck}. The tensor-to-scalar ratio is set equal to zero ($r=0$).}
\label{fig:temp_rot}
\end{figure}

We are now in the position to plot the CMB angular power spectra affected from cosmic birefringence. In order to do this we have numerically evaluated Eqs.~\eqref{eqn:EE}-\eqref{eqn:TB}, by computing each component via Eqs.~\eqref{eqn:C_EE}-\eqref{eqn:C_TB}. In order to do this, we have again exploited our modified version of \texttt{CLASS} to calculate the spectra of anisotropic birefringence and the isotropic angle from the two epochs, i.e. recombination and reionization.  Let us just mention that we have neglected all the unlensed unrotated terms coming from different sources (i.e. ``reco-reio'' and ``reio-reco''). Since we expect the CMB radiation transfer functions for the recombination and the reionization contributions to peak at very different redshifts, it is reasonable to neglect such cross-correlations. The final results are plotted in Figs.~\ref{fig:pol_rot}-\ref{fig:temp_rot} up to $\ell_{\text{max}}=200$ just for sake of simplicity, since the evaluation of Wigner $3$-$j$ symbols is numerically time-consuming. 
If we look for instance at Fig.~\ref{fig:E-E}, we observe that the $C_{\ell}^{EE}$ spectrum 
predicted by the $\Lambda$CDM model (i.e. in absence of cosmic birefringence) is larger than the rotated ones: this is due to the fact that the birefringence mechanism induces a mixing of the $E$ and $B$ polarization modes, and so it partially removes power from the former to transfer it to the latter.

By direct inspection of Figs.~\ref{fig:pol_rot}-\ref{fig:temp_rot}, we see that impact of cosmic birefringence on CMB power spectra is consistent with our previous considerations made in Sec.~\ref{sec:chi_0}. For instance, as shown in Fig.~\ref{fig:E-E}, the angular power spectrum of the $E$ modes of CMB polarization deviates from the $\Lambda$CDM one (i.e. without birefringence) for $m_{\chi}=\SI{e-31}{\electronvolt}$ at low multipoles. Indeed the dominant modification comes from isotropic cosmic birefringence, and according to Fig.~\ref{fig:chi_background} an axion field of such mass is clearly present also after reionization, causing such an impact on the low multipoles of $C^{EE}_{\ell}$. Analogously, according again to Fig.~\ref{fig:chi_background}, an axion with mass $m_{\chi}=\SI{e-28}{\electronvolt}$ does not experience a relevant time evolution after reionization: indeed, this is clearly noticed also in Fig.~\ref{fig:B-B}, where the reionization contribution to $C_{\ell}^{EE}$ at low multipoles is not significantly converted to $C^{BB}_{\ell}$. 

\subsection{ACB from Recombination and Reionization}
\begin{figure}
\centering
\subfloat[][\label{fig:AA-reio}Auto-correlation of ACB, according to Eq.~\eqref{eqn:Caa}.]
{\includegraphics[width=.75\textwidth]{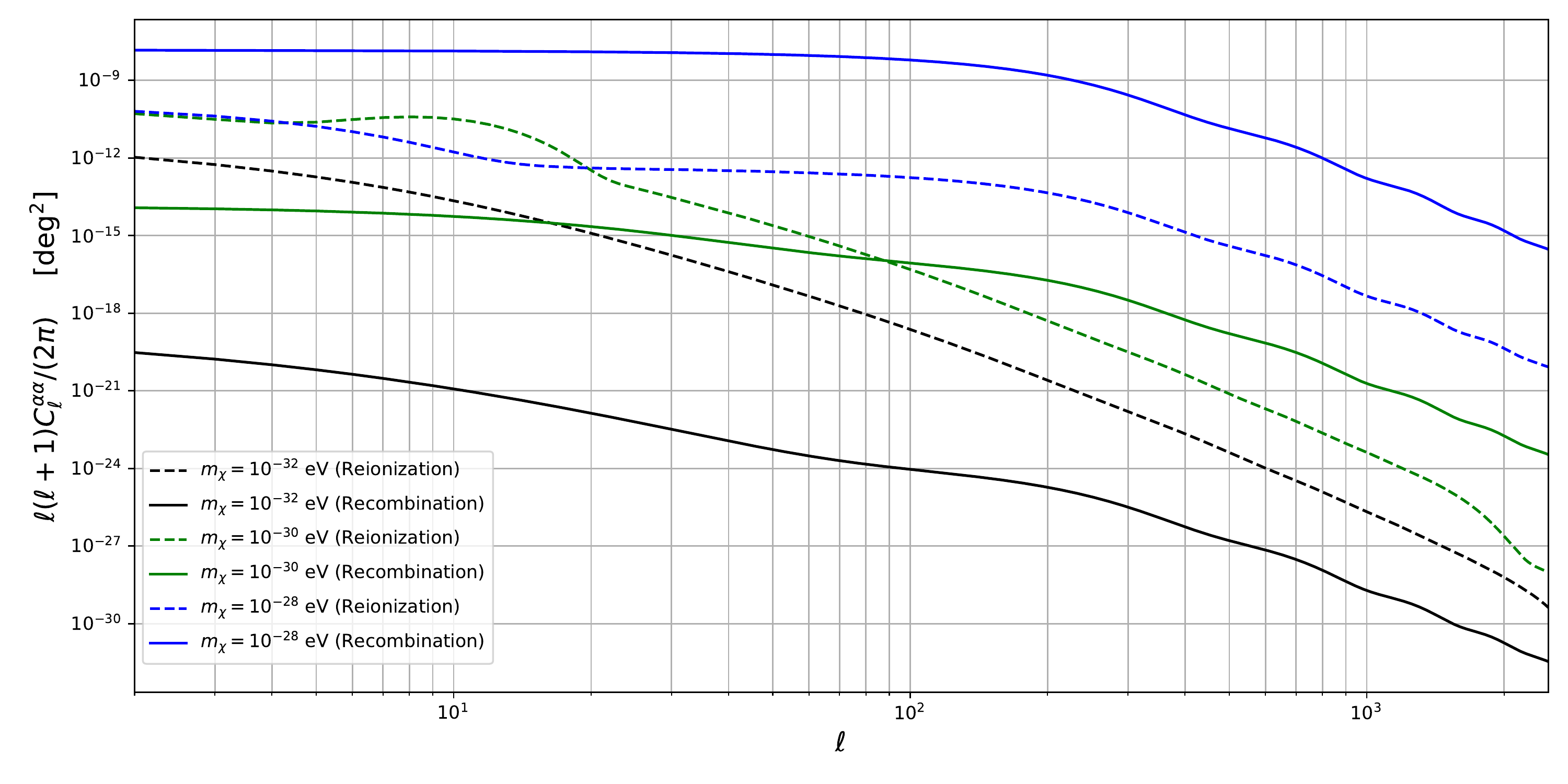}} \\
\subfloat[][\label{fig:AT-reio}Absolute value of the cross-correlation of ACB with CMB temperature, according to Eq.~\eqref{eqn:Cat}.]
{\includegraphics[width=.75\textwidth]{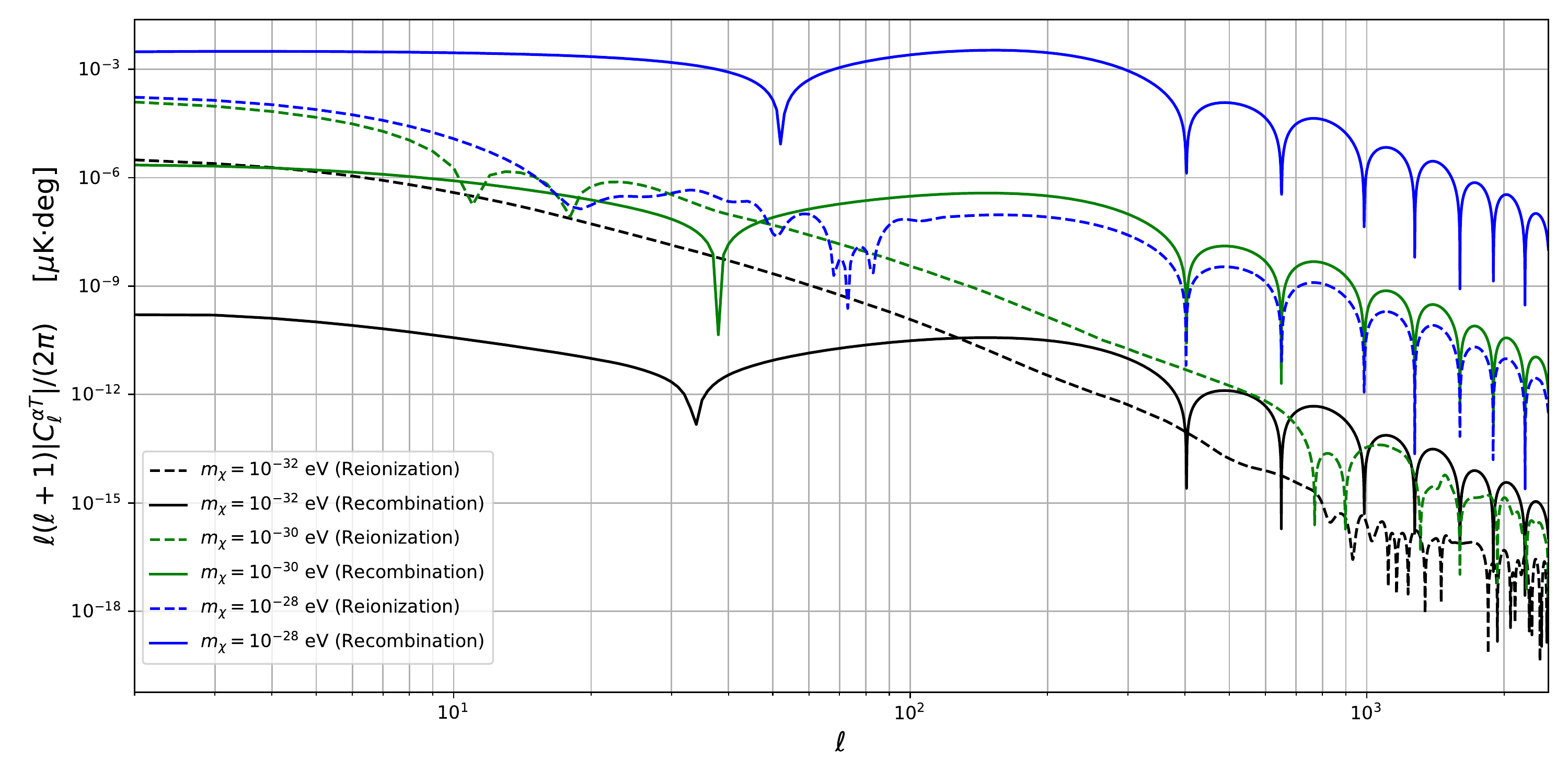}} \\
\subfloat[][\label{fig:AE-reio}Absolute value of the cross-correlation of ACB with $E$ modes of CMB polarization, according to Eq.~\eqref{eqn:Cae}.]
{\includegraphics[width=.75\textwidth]{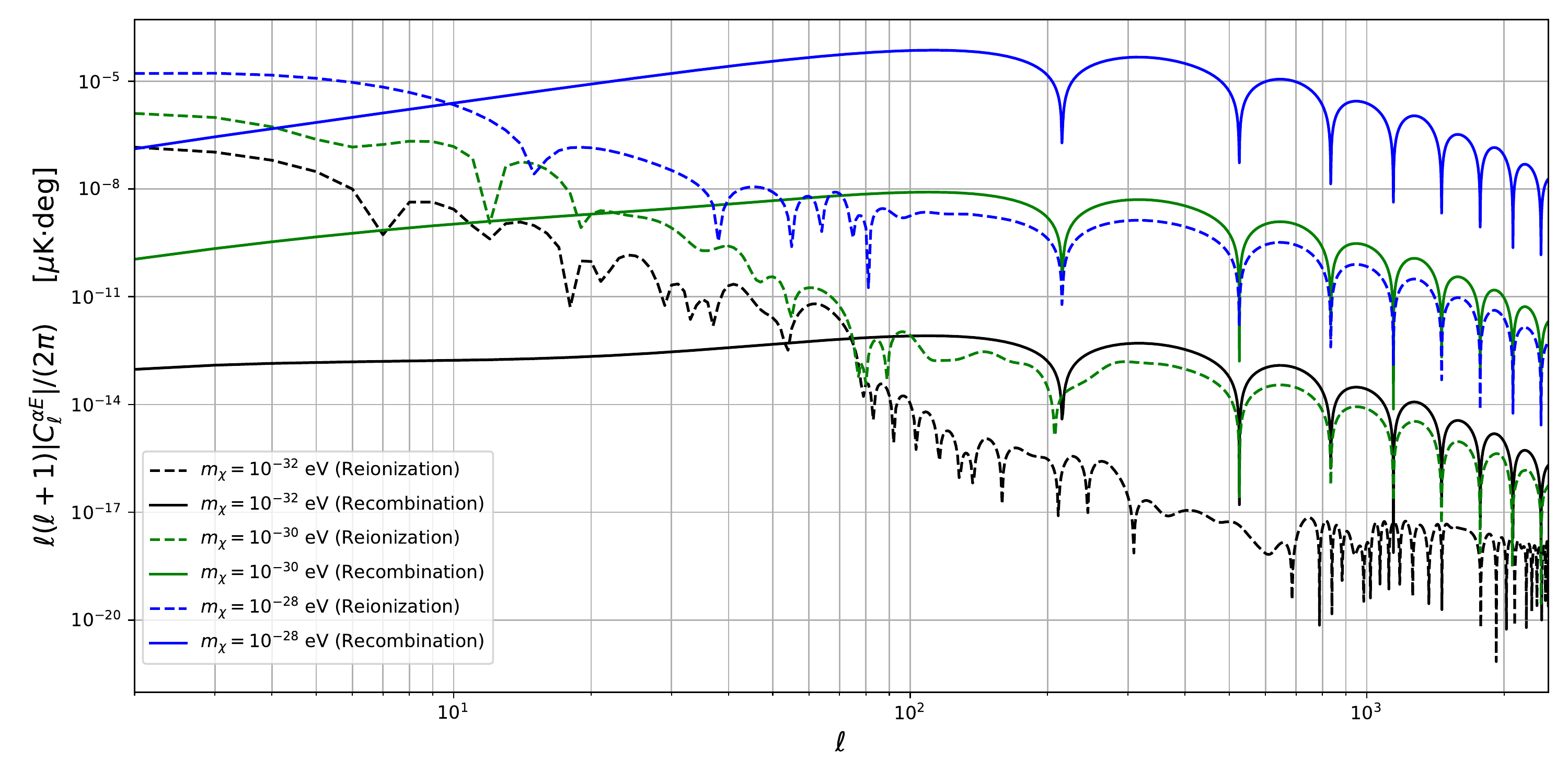}}
\caption{Angular power spectra involving anisotropic cosmic birefringence for the model defined by Eq.~\eqref{eqn:quintessence_potential}, coming from recombination (solid line) and from reionization (dashed line). The mixed terms (i.e. recombination-reionization) are not considered here. The numerical computation has been performed for several values of the field mass, by taking $\lambda/f=\SI{e-18}{\giga\electronvolt^{-1}}$, $\chi_{0}^{\text{ini}}= m_{Pl}$ (reduced Planck mass), $\chi_{0}^{\prime\,\text{ini}}=0$, and for the fiducial values of the $\Lambda$CDM parameters provided in \cite{aghanim2020planck}.}
\label{fig:reco_reio}
\end{figure}

By looking at Eqs.~\eqref{eqn:EE}-\eqref{eqn:TB}, whose components are given by the formulas collected in Eqs.~\eqref{eqn:C_EE}-\eqref{eqn:C_TB}, we can see that cosmic birefringence enters in the expressions for the CMB rotated spectra with both the recombination and reionization contributions. Thus, it is important to understand how much the two signals differ. To see this, we have used again our modified \texttt{CLASS} code to plot the angular power spectra of anisotropic birefringence from the two epochs in Fig.~\ref{fig:reco_reio}.
Let us notice that different masses of the axion field imply a different contribution to the reionization signal: indeed, it turns out that for a sufficiently light axion scalar field, the contribution from reionization can be larger than that one from recombination at least at low multipoles for the spectra we considered, i.e. $C_{\ell}^{\alpha\alpha}$, $C_{\ell}^{\alpha T}$, and $C_{\ell}^{\alpha E}$. On the other hand, as the axion mass increases, the two contributions become comparable to each other.

\section{\label{sec:End}Conclusions}
For the first time, in this paper we have performed a tomographic analysis of anisotropic cosmic birefringence, by studying the dynamics of an axion-like field $\chi$. We have solved the equation of motion of the background term $\chi_0$ and of its inhomogeneous fluctuation $\delta\chi$ for several values of the axion mass $m_{\chi}$, that enter in the axion potential as described by Eq.~\eqref{eqn:quintessence_potential}. Our approach has allowed to quantify the impact of the birefringence effects on the CMB observables. We have found that different values for the axion mass imprint very different signatures in the birefringence signal, making the tomographic approach to ACB a powerful probe of the axion field underlying physics, in analogy with what has been found in previous studies in literature, e.g. in \cite{nakatsuka2022cosmic,lee2022probing}, but where the analysis was restricted just to the isotropic case of cosmic birefringence. Indeed, we account also for the reionization contribution to anisotropic cosmic birefringence, which is something completely new up to our knowledge. 

A relevant message we want to convey is that our tomographic treatment of ACB is able to make manifest unique features of the birefringence anisotropies with respect to the purely isotropic case:  indeed, we have shown that, although a large axion mass prevents the possibility to have isotropic cosmic birefringence, this behavior is not mimicked by the anisotropic counterpart. As can be seen by comparing Fig.~\ref{fig:chi_background} with e.g. Fig.~\ref{fig:alpha-alpha}, this is due to the fact the larger the axion mass is, larger the amplitudes of the $C_{\ell}^{\alpha\alpha}$, $C_{\ell}^{\alpha T}$ and $C_{\ell}^{\alpha E}$ are. This fact has a very intriguing consequence: since CMB observations have constrained the ACB amplitude below a certain threshold, it follows that the axion mass can be constrained too up to an upper value. This is a clear example that ACB can encode additional and complementary information, relevant also for the isotropic counterpart.

Another important result we found in this paper is that for low multipoles and for sufficiently small values of the axion mass, the reionization contribution to anisotropic cosmic birefringence is higher with respect to the recombination one, as can be seen by looking at Fig.~\ref{fig:reco_reio}. For this reason, a future development of our research could be trying to use the signal coming from reionization encoded in ACB as a probe of the axion parameters, as it has been already done in just the purely isotropic regime \cite{sherwin2021cosmic,nakatsuka2022cosmic,lee2022probing}. 

All the aforementioned results were possible thanks to the generalization of the standard formalism which describes how anisotropic cosmic birefringence affects the CMB angular power spectra, by including the reionization contribution. Indeed, the general formulas collected in Eqs.~\eqref{eqn:C_EE}-\eqref{eqn:C_TB} can be seen as a generalized version of the well known equations for ACB that it is possible to find e.g. in \cite{li2008cosmological}. Thanks to our modified version of the Boltzmann code \texttt{CLASS}, we have been also able to numerically compute the rotated CMB spectra, which is one of the observables where we expect to test our theoretical predictions. A future application of our work would be a statistical forecast for the axion and birefringence parameters, including the signal coming from the reionization epoch, that it is now possible thanks to the general results of Eqs.~\eqref{eqn:C_EE}-\eqref{eqn:C_TB}.

\begin{acknowledgments}
The authors would like to thank F. Bianchini for his help in providing the $C_{\ell}^{\alpha\alpha}$ and $C_{\ell}^{\alpha T}$ spectra as observed through the South Pole Telescope \cite{bianchini2020searching}. The authors also thank the referee for carefully reading the paper and for giving constructive remarks that substantially helped improving the quality of the work. N. Bartolo and A. Gruppuso acknowledge support from the COSMOS network (www.cosmosnet.it) through the ASI (Italian Space Agency) Grants 2016-24-H.0, 2016-24-H.1-2018 and 2020-9-HH.0. 
\end{acknowledgments}

\appendix

\section{\label{App:CMB}Full Analytical Derivation of Birefringent CMB Angular Power Spectra}

In this appendix, we are going to compute all the terms collected in Eqs.~\eqref{eqn:EE}-\eqref{eqn:TB}. In order to do this, it is convenient to introduce the following compact expression for the rotated harmonic coefficients of the CMB polarization modes:
\begin{equation}
	\label{eqn:Compact_Expressions}
	\begin{split}
		a_{P_j,\ell m}^{\text{rot},x}=\sum_{s=\pm2}\frac{e^{is\alpha_0(\tau_x)}}{2}&\sum_{LM}\int\mathrm{d}\hat{n}\,_sY_{\ell m}^*(\versor{n})\,_sY_{LM}(\versor{n})\mathscr{R}^{(s)}_{jk}\,a^x_{P_k,LM}e^{is\delta\alpha(\tau_x,\versor{n})}
	\end{split}
\end{equation}
where for $j,k=1,2$ we have defined
\begin{equation}
	\label{eqn:Greek_Matrix}
	a_{P,\ell m}=\begin{pmatrix}
		a_{E,\ell m} \\
		a_{B,\ell m}
	\end{pmatrix},
	\qquad\mathscr{R}^{(s)}\equiv\begin{pmatrix}
		1 & is/2 \\
		-is/2 & 1
	\end{pmatrix},
\end{equation}
and the sum over $k$ has to be understood. 

To calculate the component of the rotated correlation functions of CMB anisotropies we
make the following assumptions:
\begin{enumerate}
	\item the anisotropic cosmic birefringence angle is small everywhere, so that we can adopt a perturbative approach;
	\item the unrotated CMB fields and the anisotropic birefringence angle can all be treated as Gaussian random fields;
	\item the underlying inflationary model is parity-conserving, so that $C_{\ell}^{EB}=0=C_{\ell}^{TB}$ for primordial unrotated modes.
\end{enumerate}
The simplest way to proceed is to evaluate the following general cross-correlator of CMB polarization:
\begin{equation}
	\label{eqn:general_correlation}
	\begin{split}
		\langle a^{\text{rot},x}_{P_j,\ell_1m_1}&a^{\text{rot},z}_{P_i,\ell_2 m_2}\rangle=\sum_{s_1s_2}\frac{e^{is_1\alpha_{0}(\tau_x)}e^{is_2\alpha_{0}(\tau_z)}}{4}\sum_{L_1M_1}\sum_{L_2M_2}\int\mathrm{d}^2\hat{n}_1\,\mathrm{d}^2\hat{n}_2\,\mathscr{R}_{jk}^{(s_1)}\mathscr{R}_{it}^{(s_2)}\\
		&\,_{s_1}Y_{\ell_1m_1}^*(\versor{n}_1)\,_{s_1}Y_{L_1M_1}(\versor{n}_1)Y_{\ell_2m_2}^*(\versor{n}_2)Y_{L_2M_2}(\versor{n}_2)\langle a^x_{P_k,L_1M_1}a^z_{P_t,L_2M_2}e^{is_1\delta\alpha(\tau_x,\versor{n}_1)}e^{is_2\delta\alpha(\tau_z,\versor{n}_2)}\rangle.
	\end{split}
\end{equation}
Thanks to our assumptions, we can Taylor-expand the exponential containing the anisotropic birefringence angle at the quadratic order:
\begin{equation}
	\label{eqn:quadratic_order}
	e^{is\delta\alpha(\tau,\versor{n})}=1+is\sum_{pq}\alpha_{pq}^{x}Y_{pq}(\versor{n})-2\sum_{pq}\sum_{uv}\alpha_{pq}^x\alpha_{uv}^{x}Y_{pq}(\versor{n})Y_{uv}(\versor{n})+\mathcal{O}(\delta\alpha^3),
\end{equation}
so that the ensemble average at the last line of Eq.~\eqref{eqn:general_correlation} can be rewritten as follows
\begin{equation}
	\label{eqn:wikkoni}
	\begin{split}
		\langle a^x_{P_k,L_1M_1}a^z_{P_t,L_2M_2}e^{is_1\delta\alpha(\tau_x,\versor{n}_1)}e^{is_2\delta\alpha(\tau_z,\versor{n}_2)}\rangle&=\expval{a^x_{P_k,L_1M_1}a^z_{P_t,L_2M_2}}\\
		&\quad-2\sum_{p_1q_1}\sum_{u_1v_1}Y_{p_1q_1}(\versor{n}_1)Y_{u_1v_1}(\versor{n}_1)\expval{a^x_{P_k,L_1M_1}a^z_{P_t,L_2M_2}\alpha^x_{p_1q_1}\alpha^x_{u_1v_1}}\\
		&\quad-2\sum_{p_2q_2}\sum_{u_2v_2}Y_{p_2q_2}(\versor{n}_2)Y_{u_2v_2}(\versor{n}_2)\expval{a^x_{P_k,L_1M_1}a^z_{P_t,L_2M_2}\alpha^z_{p_2q_2}\alpha^z_{u_2v_2}}\\
		&\quad-s_1s_2\sum_{p_1q_1}\sum_{p_2q_2}Y_{p_1q_1}(\versor{n}_1)Y_{p_2q_2}(\versor{n}_2)\expval{a^x_{P_k,L_1M_1}a^z_{P_t,L_2M_2}\alpha^x_{p_1q_1}\alpha^z_{p_2q_2}}\\
		&\quad+\mathcal{O}(\delta\alpha^3).
	\end{split}
\end{equation}
Each one of the four terms on the right-hand side of the equation above can be in turn decomposed by means of the Isserlis theorem \cite{isserlis1918formula}. For instance, the term in the second line at the right-hand side of Eq.~\eqref{eqn:wikkoni} yields
\begin{equation}
	\begin{split}
		-2&\sum_{p_1q_1}\sum_{u_1v_1}Y_{p_1q_1}(\versor{n}_1)Y_{u_1v_1}(\versor{n}_1)\expval{a^x_{P_k,L_1M_1}a^z_{P_t,L_2M_2}\alpha^x_{p_1q_1}\alpha^x_{u_1v_1}}=\\
		&=-2\sum_{p_1q_1}\sum_{u_1v_1}Y_{p_1q_1}(\versor{n}_1)Y_{u_1v_1}(\versor{n}_1)\big[\expval{a_{P_k,L_1M_1}^xa^z_{P_t,L_2M_2}}\expval{\alpha^x_{p_1q_1}\alpha^x_{u_1v_1}}+\expval{a_{P_k,L_1M_1}^x\alpha^x_{p_1q_1}}\expval{a^z_{P_t,L_2M_2}\alpha^x_{u_1v_1}}\\
		&\hspace{300pt}+\expval{a_{P_k,L_1M_1}^x\alpha^x_{u_1v_1}}\expval{a^z_{P_t,L_2M_2}\alpha^x_{p_1q_1}}\big],
	\end{split}
\end{equation}
which can be simplified with the definition of Eq.~\eqref{eqn:alpha-alpha} as
\begin{equation}
	\begin{split}
		-2&\sum_{p_1q_1}\sum_{u_1v_1}Y_{p_1q_1}(\versor{n}_1)Y_{u_1v_1}(\versor{n}_1)\expval{a^x_{P_k,L_1M_1}a^z_{P_t,L_2M_2}\alpha^x_{p_1q_1}\alpha^x_{u_1v_1}}=\\
		&=-2\sum_{p_1q_1}\sum_{u_1v_1}Y_{p_1q_1}(\versor{n}_1)Y_{u_1v_1}(\versor{n}_1)\bigg[C^{P_kP_t}_{L_1}\big|_{xz}C_{p_1}^{\alpha\alpha}\big|_{xx}\delta_{L_1L_2}\delta_{M_1,-M_2}\delta_{p_1u_1}\delta_{q_1,-v_1}\\
		&\qquad+C_{L_1}^{\alpha P_k}\big|_{xx}C_{\alpha P_t}\big|_{xz}\delta_{L_1p_1}\delta_{M_1,-q_1}\delta_{L_2,u_1}\delta_{M_2,-v_1}+C_{L_1}^{\alpha P_k}\big|_{xx}C_{L_1}^{\alpha P_t}\big|_{xz}\delta_{L_1u_1}\delta_{M_1,-v_1}\delta_{L_2p_1}\delta_{M_2,-q_1}\bigg],
	\end{split}
\end{equation}
so that it further reduces to
\begin{equation}
	\begin{split}
		-2\sum_{p_1q_1}\sum_{u_1v_1}Y_{p_1q_1}(\versor{n}_1)Y_{u_1v_1}(\versor{n}_1)&\expval{a^x_{P_k,L_1M_1}a^z_{P_t,L_2M_2}\alpha^x_{p_1q_1}\alpha^x_{u_1v_1}}=\\
		&=-2\bigg[C^{P_kP_t}_{L_1}\big|_{xz}\delta_{L_1L_2}\delta_{M_1,-M_2}\sum_{p_1q_1}Y_{p_1q_1}(\versor{n}_1)Y_{p_1q_1}^*(\versor{n}_1)C_{p_1}^{\alpha\alpha}\big|_{xx}\\
		&\qquad\qquad+\bigg(C_{L_1}^{\alpha P_k}\big|_{xx}C^{\alpha P_t}_{L_2}\big|_{xz}+C_{L_1}^{\alpha P_k}\big|_{xx}C_{L_2}^{\alpha P_t}\big|_{xz}\bigg)Y_{L_1M_1}^*(\versor{n}_1)Y_{L_2M_2}^*(\versor{n}_1)\bigg].
	\end{split}
\end{equation}
Let us notice that we can recognize the variance of the anisotropic birefringence angle within the summation in the first term at the right-hand side of the previous equation:
\begin{equation}
	\label{eqn:variance}
	\sum_{p_1q_1}Y_{p_1q_1}(\versor{n}_1)Y_{p_1q_1}^*(\versor{n}_1)C_{p_1}^{\alpha\alpha}\big|_{xx}=\sum_{p_1}\left(\frac{2p_1+1}{4\pi}\right)C^{\alpha\alpha}_{p_1}\big|_{xx}\equiv	V_{\alpha}\big|_{xx},
\end{equation}
where we have used the Unsöld's theorem \cite{unsold1927beitrage}. By repeating the same procedure for all the terms at the right-hand side of Eq.~\eqref{eqn:wikkoni}, we find after lengthy calculations
\begin{equation}
	\label{eqn:general_ensemble}
	\begin{split}
		\langle &a^x_{P_k,L_1M_1}a^z_{P_t,L_2M_2}e^{is_1\delta\alpha(\tau_x,\versor{n}_1)}e^{is_2\delta\alpha(\tau_z,\versor{n}_2)}\rangle=\\
		&\quad=C^{P_kP_t}_{L_1}\big|_{xz}\left(1-2V_{\alpha}\big|_{xx}-2V_{\alpha}\big|_{zz}\right)\delta_{L_1L_2}\delta_{M_1,-M_2}-4C^{\alpha P_k}_{L_1}\big|_{xx}C^{\alpha P_t}_{L_2}\big|_{xz}Y_{L_1M_1}^*(\versor{n}_1)Y_{L_2M_2}^*(\versor{n}_1)\\
		&\qquad-4C^{\alpha P_k}_{L_1}\big|_{xz}C^{\alpha P_t}_{L_2}\big|_{zz}Y_{L_1M_1}^*(\versor{n}_2)Y_{L_2M_2}^*(\versor{n}_2)-s_1s_2\sum_{p_1q_1}C^{P_kP_t}_{L_1}\big|_{xz}C^{\alpha\alpha}_{p_1}\big|_{xz}Y_{p_1q_1}^*(\versor{n}_1)Y_{p_1q_1}^*(\versor{n}_2)\delta_{L_1L_2}\delta_{M_1,-M_2}\\
		&\qquad-s_1s_2C^{\alpha P_k}_{L_1}\big|_{xx}C^{\alpha P_t}_{L_2}\big|_{zz}Y_{L_1M_1}^*(\versor{n}_1)Y^*_{L_2M_2}(\versor{n}_2)-s_1s_2C^{\alpha P_k}_{L_1}\big|_{xz}C^{\alpha P_t}_{L_2}\big|_{xz}Y_{L_2M_2}^*(\versor{n}_1)Y_{L_1M_1}^*(\versor{n}_2).
	\end{split}
\end{equation}
Now we substitute Eq.~\eqref{eqn:general_ensemble} within Eq.~\eqref{eqn:general_correlation}, so that at the very end we have just to compute the following object:
\begin{equation}
	\label{eqn:brief_form}
	\langle a^{\text{rot},x}_{P_j,\ell_1m_1}a^{\text{rot},z}_{P_i,\ell_2 m_2}\rangle=\frac{1}{4}\sum_{s_1s_2}e^{is_1\alpha_{0}(\tau_x)}e^{is_2\alpha_{0}(\tau_z)}\mathscr{R}_{jk}^{(s_1)}\mathscr{R}_{it}^{(s_2)}\left[\text{I}+\text{II}+\text{III}+\text{IV}+\text{V}+\text{VI}\right],
\end{equation}
with
\begin{equation}
	\begin{split}
		\text{I}&\equiv\left(1-2V_{\alpha}\big|_{xx}-2V_{\alpha}\big|_{zz}\right)\sum_{L_1M_1}\sum_{L_2M_2}C_{L_1}^{P_kP_t}\big|_{xz}\delta_{L_1L_2}\delta_{M_1,-M_2}\int\mathrm{d}^2\hat{n}_1\,,_{s_1}Y^*_{\ell_1m_1}(\versor{n}_1)\,_{s_1}Y_{L_1M_1}(\versor{n}_1)\\
		&\hspace{270pt}\int\mathrm{d}^2\hat{n}\,{s_2}Y^*_{\ell_2m_2}(\versor{n}_2)\,_{s_2}Y_{L_2M_2}(\versor{n}_2)\\
		&=\left(1-2V_{\alpha}\big|_{xx}-2V_{\alpha}\big|_{zz}\right)C_{L_1}^{P_kP_t}\big|_{xz}\delta_{\ell_1\ell_2}\delta_{m_1,-m_2}.
	\end{split}
\end{equation}
where we have used Eq.~\eqref{eqn:orthonormality} to perform the angular integration. Let us proceed in the evaluation of all the pieces in Eq.~\eqref{eqn:brief_form}: the second term is
\begin{equation}
	\begin{split}
		\text{II}&\equiv-4\sum_{L_1M_1}\sum_{L_2M_2}C^{\alpha P_k}_{L_1}\big|_{zx}C^{\alpha P_t}_{L_2}\big|_{zz}\int\mathrm{d}^2\hat{n}_1\,_{s_1}Y^*_{\ell_1m_1}(\versor{n}_1)\,_{s_1}Y_{L_1M_1}(\versor{n}_1)\\
		&\hspace{150pt}\int\mathrm{d}^2\hat{n}_2\,_{s_2}Y^*_{\ell_2m_2}(\versor{n}_2)\,_{s_2}Y_{L_2M_2}(\versor{n}_2)Y_{L_1M_1}^*(\versor{n}_2)Y_{L_2M_2}^*(\versor{n}_2)\\
		&=-4\sum_{L_2M_2}C^{\alpha P_k}_{\ell_1}\big|_{zx}\sum_{L_2}C^{\alpha P_t}_{L_2}\big|_{zz}\int\mathrm{d}^2\hat{n}_2\,_{s_2}Y^*_{\ell_2m_2}(\versor{n}_2)Y_{\ell_1m_1}^*(\versor{n}_2)\sum_{M_2=-L_2}^{L_2}\,_{s_2}Y_{L_2M_2}(\versor{n}_2)Y_{L_2M_2}^*(\versor{n}_2)\\
		&=0
	\end{split}
\end{equation}
because of the generalized addition theorem of spin-weighted spherical harmonics \cite{goldberg1967spin}:
\begin{equation}
	\label{eqn:vanishing_terms}
	\sum_{m=-\ell}^{\ell}\,_{s}Y_{\ell m}(\versor{n})\,_{s'}Y_{\ell m}^*(\versor{n})=\frac{2\ell+1}{4\pi}\delta_{ss'}.
\end{equation}
For the same reason, also the third term identically vanishes
\begin{equation}
	\begin{split}
		\text{III}&\equiv-4\sum_{L_1M_1}\sum_{L_2M_2}C^{\alpha P_k}_{L_1}\big|_{xx}C^{\alpha P_t}_{L_2}\big|_{xz}\int\mathrm{d}^2\hat{n}_1\,_{s_1}Y^*_{\ell_1m_1}(\versor{n}_1)\,_{s_1}Y_{L_1M_1}(\versor{n}_1)Y_{L_1M_1}^*(\versor{n}_1)Y_{L_2M_2}^*(\versor{n}_1)\\
		&\hspace{250pt}\int\mathrm{d}^2\hat{n}_2\,_{s_2}Y^*_{\ell_2m_2}(\versor{n}_2)\,_{s_2}Y_{L_2M_2}(\versor{n}_2)\\
		&=-4\sum_{L_1}C^{\alpha P_k}_{L_1}\big|_{xx}C^{\alpha P_t}_{L_2}\big|_{xz}\int\mathrm{d}^2\hat{n}_1\,_{s_1}Y^*_{\ell_1m_1}(\versor{n}_1)Y_{\ell_2m_2}^*(\versor{n}_1)\sum_{M_2=-L_2}^{L_2}\,_{s_1}Y_{L_1M_1}(\versor{n}_1)Y_{L_1M_1}^*(\versor{n}_1)\\
		&=0.\\
	\end{split}
\end{equation}
Evaluating the fourth term is more tricky:
\begin{equation}
	\label{eqn:fourth}
	\begin{split}
		\text{IV}&\equiv-s_1s_2\sum_{L_1M_1}\sum_{L_2M_2}\sum_{p_1q_1}C^{P_kP_t}_{L_1}\big|_{xz}C^{\alpha\alpha}_{p_1}\big|_{xz}\delta_{L_1L_2}\delta_{M_1,-M_2}\\
		&\hspace{50pt}\int\mathrm{d}^2\hat{n}_1\,_{s_1}Y_{\ell_1m_1}^*(\versor{n}_1)\,_{s_1}Y_{L_1M_1}(\versor{n}_2)Y_{p_1q_1}(\versor{n}_1)\int\mathrm{d}^2\hat{n}_2\,_{s_2}Y_{\ell_2m_2}^*(\versor{n}_2)\,_{s_2}Y_{L_2M_2}(\versor{n}_2)Y_{p_1q_1}^*(\versor{n}_2),
	\end{split}
\end{equation}
since, in order perform the angular integration, we have to exploit a formula for the triple integral \cite{newman1966note}:
\begin{equation}
	\label{eqn:triple}
	\begin{split}
		\int\mathrm{d}^2\hat{n}&\,_{s_1}Y_{\ell_1 ,m_1}(\versor{n})\,_{s_2}Y_{\ell_2,m_2}(\versor{n})\,_{s_3}Y_{\ell_3,m_3}(\versor{n})=I^{-s_1,-s_2,-s_3}_{\ell_1\ell_2\ell_3}\begin{pmatrix}
			\ell_1 & \ell_2 & \ell_3 \\
			m_1 & m_2 & m_3
		\end{pmatrix},
	\end{split}
\end{equation}
where we have defined
\begin{equation}
	\begin{split}
		I^{-s_1,-s_2,-s_3}_{\ell_1\ell_2\ell_3}&\equiv\sqrt{\frac{(2\ell_1+1)(2\ell_2+1)(2\ell_3+1)}{4\pi}}\begin{pmatrix}
			\ell_1 & \ell_2 & \ell_3 \\
			-s_1 & -s_2 & -s_3
		\end{pmatrix},
	\end{split}
\end{equation}
and where the ``matrix'' is a  Wigner $3j$-symbol, which satisfies the selection rule and triangle conditions \cite{varshalovich1988quantum},
\begin{equation}
	|m_i|\le\ell_i\,\forall i=1,2,3,\qquad m_1+m_2=m_3,\qquad|\ell_1-\ell_2|\le\ell_3\le|\ell_1+\ell_2|,
\end{equation}
and obeys the symmetry:
\begin{equation}
	\begin{split}
		\begin{pmatrix}
			\ell_1 & \ell_2 & \ell_3 \\
			m_1 & m_2 & m_3
		\end{pmatrix}&=\begin{pmatrix}
			\ell_2 & \ell_3 & \ell_1 \\
			m_2 & m_3 & m_1
		\end{pmatrix}=\begin{pmatrix}
			\ell_3 & \ell_1 & \ell_2 \\
			m_3 & m_1 & m_2
		\end{pmatrix}\\
		&=(-1)^{\ell_T}\begin{pmatrix}
			\ell_1 & \ell_3 & \ell_2 \\
			m_1 & m_3 & m_2
		\end{pmatrix}=(-1)^{\ell_T}\begin{pmatrix}
			\ell_3 & \ell_2 & \ell_1 \\
			m_3 & m_2 & m_1
		\end{pmatrix}=(-1)^{\ell_T}\begin{pmatrix}
			\ell_2 & \ell_1 & \ell_3 \\
			m_2 & m_1 & m_3
		\end{pmatrix}\\
		&=(-1)^{\ell_T}\begin{pmatrix}
			\ell_1 & \ell_2 & \ell_3 \\
			-m_1 & -m_2 & -m_3
		\end{pmatrix},
	\end{split}
\end{equation}
$\ell_T\equiv\ell_1+\ell_2+\ell_3$ being the total multipoles number. Therefore, Eq.~\eqref{eqn:fourth} can be rewritten as
\begin{equation}
	\begin{split}
		\text{IV}&=-s_1s_2\sum_{L_1M_1}\sum_{p_1q_1}C^{P_kP_t}_{L_1}\big|_{xz}C^{\alpha\alpha}_{p_1}\big|_{xz}I^{s_1,-s_1,0}_{\ell_1L_1p_1}I^{s_2,-s_2,0}_{\ell_2L_1p_1}\begin{pmatrix}
			\ell_1 & L_1 & p_1 \\
			-m_1 & M_1 & q_1
		\end{pmatrix}
		\begin{pmatrix}
			\ell_2 & L_1 & p_1 \\
			-m_2 & -M_2 & -q_1
		\end{pmatrix}\\
		&=-\frac{s_1s_2}{2\ell_1+1}\delta_{\ell_1\ell_2}\delta_{m_1,-m_2}\sum_{L_1p_1}(-1)^{\ell_1+L_1+p_1}C_{L_1}^{P_kP_t}\big|_{xz}C^{\alpha\alpha}_{p_1}\big|_{xz}I^{s_1,s_1,0}_{\ell_1L_1p_1}I_{\ell_2L_1p_1}^{s_2,-s_2,0}
	\end{split}
\end{equation}
where we have used the orthogonality relation of the Wigner 3$j$-symbol \cite{varshalovich1988quantum}:
\begin{equation}
	\sum_{m_1m_2}\begin{pmatrix}
		\ell_1 & \ell_2 & \ell_3\\
		m_1 & m_2 & m_3
	\end{pmatrix}\begin{pmatrix}
		\ell_1 & \ell_2 & \ell_3'\\
		m_1 & m_2 & m_3'
	\end{pmatrix}=\frac{\delta_{\ell_3\ell_3'}\delta_{m_3m_3'}}{2\ell_3+1}.
\end{equation}
At the contrary, we now show that the fifth term vanishes:
\begin{equation}
	\begin{split}
		\text{V}&\equiv-s_1s_2\sum_{L_1M_1}\sum_{L_2M_2}C^{\alpha P_k}_{L_1}\big|_{xx}C^{\alpha P_t}_{L_2}\big|_{zz}\int\mathrm{d}^2\hat{n}_1\,_{s_1}Y_{\ell_1m_1}^*(\versor{n}_1)\,_{s_1}Y_{L_1M_1}(\versor{n}_1)Y_{L_1M_1}^*(\versor{n}_1)\\
		&\hspace{200pt}\int\mathrm{d}^2\hat{n}_2\,_{s_2}Y_{\ell_2m_2}^*(\versor{n}_2)\,_{s_2}Y_{L_2M_2}(\versor{n}_2)Y_{L_2M_2}^*(\versor{n}_2)\\
		&=-s_1s_2\sum_{L_1M_1}\sum_{L_2M_2}C^{\alpha P_k}_{L_1}\big|_{xx}C^{\alpha P_t}_{L_2}\big|_{zz}I^{-s_1,s_1,0}_{\ell_1 L_1 L_1}I_{\ell_2L_2L_2}^{-s_2,s_2,0}\begin{pmatrix}
			\ell_1 & L_1 & L_1 \\
			-m_1 & M_1 & -M_1
		\end{pmatrix}\begin{pmatrix}
			\ell_2 & L_2 & L_2\\
			-m_2 & M_2 & -M_2
		\end{pmatrix}\\
		&=0.
	\end{split}
\end{equation}
Indeed, the expression above can be set equal to zero because the selection rule of the Wigner $3j$-symbol forces $m_1=m_2=0$, but this implies e.g. \cite{landau2013quantum}
\begin{equation}
	\begin{pmatrix}
		\ell_1 & L_1 &L_1 \\
		0 & M_1 & -1M_1
	\end{pmatrix}\propto\delta_{\ell_10}=0,
\end{equation}
since the monopole $\ell_1=0$ is not observable. Finally the last term is given as
\begin{equation}
	\begin{split}
		\text{VI}&\equiv-s_1s_2\sum_{L_1M_1}\sum_{L_2M_2}C^{\alpha P_k}_{L_1}\big|_{zx}C^{\alpha P_t}_{L_2}\big|_{xz}\int\mathrm{d}^2\hat{n}_1\,_{s_1}Y_{\ell_1m_1}^*(\versor{n}_1)\,_{s_1}Y_{L_1M_1}(\versor{n}_1)Y_{L_2M_2}^*(\versor{n}_1)\\
		&\hspace{175pt}\int\mathrm{d}^2\hat{n}_2\,_{s_2}Y_{\ell_2m_2}^*(\versor{n}_2)\,_{s_2}Y_{L_2M_2}(\versor{n}_2)Y_{L_1M_1}^*(\versor{n}_2)\\
		&=-s_1s_2\sum_{L_1M_1}\sum_{L_2M_2}(-1)^{\ell_2+L_1+L_2}C^{\alpha P_k}_{L_1}\big|_{zx}C^{\alpha P_t}_{L_2}\big|_{xz}I_{\ell_1L_1L_2}^{s_1,-s_1,0}I^{s_2,0,-s_2}_{\ell_2L_1L_2}\begin{pmatrix}
			\ell_1 & L_1 & L_2 \\
			-m_1 & M_1 & -M_2
		\end{pmatrix}\begin{pmatrix}
			\ell_2 & L_1 & L_2 \\
			m_2 & M_1 & -M_2
		\end{pmatrix}\\
		&=-\frac{s_1s_2}{2\ell_1+1}\delta_{\ell_1\ell_2}\delta_{m_1,-m_2}\sum_{L_1L_2}C^{\alpha P_k}_{L_1}\big|_{zx}C^{\alpha P_t}_{L_2}\big|_{xz}I_{\ell_1L_1L_2}^{s_1,-s_1,0}I^{s_2,0,-s_2}_{\ell_1L_1L_2}.
	\end{split}
\end{equation}
We are now in the position to plug all the terms in the right-hand side of Eq.~\eqref{eqn:brief_form} together and find the following general formula:
\begin{equation}
\label{eqn:final_formula}
\begin{split}
C_{\ell,\text{rot}}^{P_jP_i}\big|_{xz}=\frac{1}{4}&\sum_{s_1s_2}e^{is_1\alpha_{0}(\tau_x)}e^{is_2\alpha_0(\tau_z)}\mathscr{R}_{jk}^{(s_1)}\mathscr{R}_{it}^{(s_2)}\Bigg[\left(1-2V_{\alpha}\big|_{xx}-2V_{\alpha}\big|_{zz}\right)C_{\ell}^{P_kP_t}\big|_{xz}\\
&-\frac{s_1s_2}{2\ell+1}\sum_{L_1L_2}(-1)^{\ell+L_1+L_2}I_{\ell L_1L_2}^{s_1,-s_1,0}\left(C_{L_1}^{P_kP_t}\big|_{xz}C^{\alpha\alpha}_{L_2}\big|_{xz}I^{s_2,-s_2,0}_{\ell L_1 L_2}+C^{\alpha P_k}_{L_1}\big|_{zx}C^{\alpha P_t}_{L_2}\big|_{xz}I_{\ell L_1 L_2}^{s_2,0,-s_2}\right)\Bigg].
\end{split}
\end{equation}

The expression above has to be specialized for the cases of interest: this is done by performing the summation over $s_1,s_2=\pm2$ for the elements of the vector $a_{P,\ell m}$ defined in Eq.~\eqref{eqn:Greek_Matrix}:
\begin{equation}
\label{eqn:APP_EE}
\begin{split}
&C_{\ell,\text{rot}}^{EE}\big|_{xz}=\left(1-2V_{\alpha}\big|_{xx}-2V_{\alpha}\big|_{zz}\right)\left[C^{EE}_{\ell}\big|_{xz}\cos(2\alpha_{0,x})\cos(2\alpha_{0,z})+C_{\ell}^{BB}\big|_{xz}\sin(2\alpha_{0,x})\sin(2\alpha_{0,z})\right]\\
&\quad+\frac{2}{2\ell+1}\sum_{L_1L_2}I^{2,-2,0}_{\ell L_1L_2}\Big(C^{\alpha\alpha}_{L_2}\big|_{xz}I^{2,-2,0}_{\ell L_1L_2}\Big\{C^{EE}_{L_1}\big|_{xz}\left[\cos(2\alpha_{0,x}-2\alpha_{0,z})-(-1)^{\ell+L_1+L_2}\cos(2\alpha_{0,x}+2\alpha_{0,z})\right]\\
&\hspace{170pt}+C^{BB}_{L_1}\big|_{xz}\left[\cos(2\alpha_{0,x}-2\alpha_{0,z})+(-1)^{\ell+ L_1+L_2}\cos(2\alpha_{0,x}+2\alpha_{0,z})\right]\Big\}\\
&\hspace{120pt}+C^{\alpha E}_{L_1}\big|_{xz}C^{\alpha E}_{L_2}\big|_{zx}I^{2,0,-2}_{\ell L_1L_2}\left[\cos(2\alpha_{0,x}-2\alpha_{0,z})-(-1)^{\ell+L_1+L_2}\cos(2\alpha_{0,x}+2\alpha_{0,z})\right]\Big),
\end{split}
\end{equation}
\begin{equation}
\label{eqn:APP_BB}
\begin{split}
&C_{\ell,\text{rot}}^{BB}\big|_{xz}=\left(1-2V_{\alpha}\big|_{xx}-2V_{\alpha}\big|_{zz}\right)\left[C^{BB}_{\ell}\big|_{xz}\cos(2\alpha_{0,x})\cos(2\alpha_{0,z})+C_{\ell}^{EE}\big|_{xz}\sin(2\alpha_{0,x})\sin(2\alpha_{0,z})\right]\\
&\quad+\frac{2}{2\ell+1}\sum_{L_1L_2}I^{2,-2,0}_{\ell L_1L_2}\Big(C^{\alpha\alpha}_{L_2}\big|_{xz}I^{2,-2,0}_{\ell L_1L_2}\Big\{C^{EE}_{L_1}\big|_{xz}\left[\cos(2\alpha_{0,x}-2\alpha_{0,z})+(-1)^{\ell+L_1+L_2}\cos(2\alpha_{0,x}+2\alpha_{0,z})\right]\\
&\hspace{170pt}+C^{BB}_{L_1}\big|_{xz}\left[\cos(2\alpha_{0,x}-2\alpha_{0,z})-(-1)^{\ell+ L_1+L_2}\cos(2\alpha_{0,x}+2\alpha_{0,z})\right]\Big\}\\
&\hspace{120pt}+C^{\alpha E}_{L_1}\big|_{xz}C^{\alpha E}_{L_2}\big|_{zx}I^{2,0,-2}_{\ell L_1L_2}\left[\cos(2\alpha_{0,x}-2\alpha_{0,z})+(-1)^{\ell+L_1+L_2}\cos(2\alpha_{0,x}+2\alpha_{0,z})\right]\Big),
\end{split}
\end{equation}
\begin{equation}
\label{eqn:APP_EB}
\begin{split}
&C_{\ell,\text{rot}}^{EB}\big|_{xz}=\left(1-2V_{\alpha}\big|_{xx}-2V_{\alpha}\big|_{zz}\right)\left[C^{EE}_{\ell}\big|_{xz}\cos(2\alpha_{0,x})\sin(2\alpha_{0,z})-C_{\ell}^{BB}\big|_{xz}\sin(2\alpha_{0,x})\cos(2\alpha_{0,z})\right]\\
&\quad+\frac{2}{2\ell+1}\sum_{L_1L_2}I^{2,-2,0}_{\ell L_1L_2}\Big(C^{\alpha\alpha}_{L_2}\big|_{xz}I^{2,-2,0}_{\ell L_1L_2}\Big\{C^{BB}_{L_1}\big|_{xz}\left[\sin(2\alpha_{0,x}-2\alpha_{0,z})-(-1)^{\ell+L_1+L_2}\sin(2\alpha_{0,x}+2\alpha_{0,z})\right]\\
&\hspace{170pt}-C^{EE}_{L_1}\big|_{xz}\left[\sin(2\alpha_{0,x}-2\alpha_{0,z})+(-1)^{\ell+ L_1+L_2}\sin(2\alpha_{0,x}+2\alpha_{0,z})\right]\Big\}\\
&\hspace{120pt}-C^{\alpha E}_{L_1}\big|_{xz}C^{\alpha E}_{L_2}\big|_{zx}I^{2,0,-2}_{\ell L_1L_2}\left[\sin(2\alpha_{0,x}-2\alpha_{0,z})-(-1)^{\ell+L_1+L_2}\sin(2\alpha_{0,x}+2\alpha_{0,z})\right]\Big),
\end{split}
\end{equation}
where we have introduced the short-hand notation $\alpha_{0}(\tau_x)\equiv\alpha_{0,x}$.

For completeness, it is then possible to find similar formulas also for the CMB angular power spectra involving a single polarization field, i.e. $C_{\ell,\text{rot}}^{TE}$ and $C_{\ell,\text{rot}}^{TB}$. The procedure is the same previously described: one has to start from the following general cross-correlator:
\begin{equation}
	\langle a_{T,\ell_1m_1}^xa^{\text{rot},z}_{P_j,\ell_2 m_2}\rangle=\sum_{s=\pm2}\frac{e^{si\alpha_{0}(\tau_z)}}{2}\sum_{LM}\int\mathrm{d}^2\hat{n}\,_{s}Y^{*}_{\ell_2m_2}(\versor{n}_2)_{s}Y_{LM}(\versor{n})\mathscr{R}_{jk}^{(s)} \langle a_{T,\ell_1m_1}^xa^z_{P_k,LM}e^{is\delta\alpha(\tau_z,\versor{n})}\rangle,
\end{equation}  
and then expand the exponential containing the anisotropic cosmic birefringence angle at the quadratic order with Eq.~\eqref{eqn:quadratic_order}. By using again the Isserlis theorem, we can unpack the resulting four-point correlation function in terms of angular power spectra, and the final result can be simplified by exploiting Eqs.~\eqref{eqn:variance} and~\eqref{eqn:vanishing_terms}. At the very end, we find
\begin{equation}
	C_{\ell,\text{rot}}^{TP_j}\big|_{xz}=\left(1-2V_{\alpha}\big|_{zz}\right)\sum_{s=\pm2}\frac{e^{is\alpha_{0}(\tau_z)}}{2}\mathscr{R}_{jk}^{(s)}C_{\ell}^{TP_k}\big|_{xz},
\end{equation}
which yields
\begin{align}
	\label{eqn:APP_TE}
	C^{TE}_{\ell,\text{rot}}\big|_{xz}&=\left(1-2V_{\alpha}\big|_{zz}\right)\cos(2\alpha_{0,z})C^{TE}_{\ell}\big|_{xz},\\
	\label{eqn:APP_TB}
	C^{TB}_{\ell,\text{rot}}\big|_{xz}&=\left(1-2V_{\alpha}\big|_{zz}\right)\sin(2\alpha_{0,z})C^{TB}_{\ell}\big|_{xz}.
\end{align}

By setting $x=y=\text{reco}$, we can find some useful formulas, valid when only the contribution coming from recombination is considered, i.e.
\begin{equation}
\label{eqn:reco_EE}
\begin{split}
&C_{\ell,\text{rot}}^{EE}=\left(1-4V_{\alpha}\right)\left[C^{EE}_{\ell}\cos^2(2\alpha_{0})+C_{\ell}^{BB}\sin^2(2\alpha_{0})\right]\\
&\quad+\frac{2}{2\ell+1}\sum_{L_1L_2}I^{2,-2,0}_{\ell L_1L_2}\Big(I^{2,-2,0}_{\ell L_1L_2}\Big\{C^{EE}_{L_1}\left[1-(-1)^{\ell+L_1+L_2}\cos(4\alpha_{0})\right]+C^{BB}_{\ell}\left[1+(-1)^{\ell+L_1+L_2}\cos(4\alpha_{0})\right]\Big\}C_{L_2}^{\alpha\alpha}\\
&\hspace{110pt}+I^{2,0,-2}_{\ell L_1L_2}C^{\alpha E}_{L_1}\left[1-(-1)^{\ell+L_1+L_2}\cos(4\alpha_{0})\right]C^{\alpha E}_{L_2}\Big),
\end{split}
\end{equation}
\begin{equation}
\begin{split}
&C_{\ell,\text{rot}}^{BB}=\left(1-4V_{\alpha}\right)\left[C^{BB}_{\ell}\cos^2(2\alpha_{0})+C_{\ell}^{EE}\sin^2(2\alpha_{0})\right]\\
&\quad+\frac{2}{2\ell+1}\sum_{L_1L_2}I^{2,-2,0}_{\ell L_1L_2}\Big(I^{2,-2,0}_{\ell L_1L_2}\Big\{C^{EE}_{L_1}\left[1+(-1)^{\ell+L_1+L_2}\cos(4\alpha_{0})\right]+C^{BB}_{\ell}\left[1-(-1)^{\ell+L_1+L_2}\cos(4\alpha_{0})\right]\Big\}C_{L_2}^{\alpha\alpha}\\
&\hspace{110pt}+I^{2,0,-2}_{\ell L_1L_2}C^{\alpha E}_{L_1}\left[1+(-1)^{\ell+L_1+L_2}\cos(4\alpha_{0})\right]C^{\alpha E}_{L_2}\Big),
\end{split}
\end{equation}
\begin{equation}
\begin{split}
&C_{\ell,\text{rot}}^{EB}=\Bigg\{	\frac{1}{2}\left(1-4V_{\alpha}\right)\left[C^{EE}_{\ell}-C_{\ell}^{BB}\right]\\
&\hspace{50pt}-\frac{2}{2\ell+1}\sum_{L_1L_2}(-1)^{\ell+L_1+L_1}I^{2,-2,0}_{\ell L_1L_2}\left[I^{2,-2,0}_{\ell L_1L_2}\left(C^{EE}_{L_1}+C^{BB}_{L_1}\right)C_{L_2}^{\alpha\alpha}-I^{2,0,-2}_{\ell L_1L_2}C^{\alpha E}_{L_1}C^{\alpha E}_{L_2}\right]\Bigg\}\sin(4\alpha_{0}),
\end{split}
\end{equation}
\begin{equation}
C^{TE}_{\ell,\text{rot}}=\left(1-2V_{\alpha}\right)\cos(2\alpha_{0})C^{TE}_{\ell},\hspace{310pt}
\end{equation}
\begin{equation}
\label{eqn:reco_TB}
C^{TB}_{\ell,\text{rot}}=\left(1-2V_{\alpha}\right)\sin(2\alpha_{0})C^{TE}_{\ell},\hspace{310pt}
\end{equation}
where we have omitted the subscript ``reco'', since the reionization signal is not taken into account, and so the fact that the above expression is related to the recombination signal has to be understood. Moreover, let use notice that if we also disregard the cross-correlation $C_{\ell}^{\alpha E}$, then Eqs.~\eqref{eqn:reco_EE}-\eqref{eqn:reco_TB} just reduce to the standard formulas that can be found e.g. in \cite{li2008cosmological}.

\nocite{*}
\bibliography{AniTomo.bib}
	
\end{document}